\documentclass[sigconf]{acmart}

\renewcommand{\textrightarrow}{$\rightarrow$}
\usepackage{flushend}
\usepackage{xcolor}
\usepackage{booktabs}
\usepackage{graphicx}
\usepackage{paralist}
\usepackage{subfigure}
\usepackage[hang,flushmargin]{footmisc}
\renewcommand{\footnotesize}{\fontsize{8}{9}\selectfont}

\definecolor{linkcol}{rgb}{0,0,0.5}
\definecolor{citecol}{rgb}{0,0.5,0.3}
\definecolor{urlcol}{rgb}{0.3,0,0}

\usepackage{xspace}

\newif\ifcomment
\commenttrue
\ifcomment
\newcommand{\XXX}[2]{{\bf \textcolor{blue}{#1: #2}}}
\newcommand{\jbnote}[1]{{\bf \textcolor{magenta}{JB: #1}}}
\newcommand{\msnote}[1]{{\bf \textcolor{magenta}{MS: #1}}}
\newcommand{\il}[1]{{\bf \textcolor{blue}{IL: #1}}}
\newcommand{\gs}[1]{{\bf \textcolor{red}{gs: #1}}}
\newcommand{\tc}[1]{{\bf \textcolor{orange}{TC: #1}}}
\else
\newcommand{\XXX}[2]{}
\newcommand{\jbnote}[1]{}
\newcommand{\msnote}[1]{}
\newcommand{\il}[1]{}
\newcommand{\gs}[1]{}
\newcommand{\tc}[1]{}
\fi

\newcommand{\descr}[1]{\smallskip\noindent\textbf{#1}}

\newcommand{\dspol}{{\sf{\fontsize{9}{10}\selectfont /pol/}}\xspace}
\newcommand{\dssp}{{\sf{\fontsize{9}{10}\selectfont /sp/}}\xspace}
\newcommand{\dsint}{{\sf{\fontsize{9}{10}\selectfont /int/}}\xspace}
\newcommand{\dssci}{{\sf{\fontsize{9}{10}\selectfont /sci/}}\xspace}

\usepackage{url}
\makeatletter
\def\url@leostyle{%
  \@ifundefined{selectfont}{\def\UrlFont{}}%
  {\def\UrlFont{}}%
}
\makeatother
\usepackage[hyphenbreaks]{breakurl}

\definecolor{darkred}{RGB}{153,0,0}
\definecolor{darkblue}{RGB}{0,0,119}
\hypersetup{colorlinks=true, linkcolor = darkred,   citecolor = darkred, urlcolor = darkblue}

\renewcommand\footnotetextcopyrightpermission[1]{} %
\setcopyright{none} %

\begin{document}

\title{The Web Centipede: Understanding How Web Communities Influence Each Other Through the Lens of Mainstream and Alternative News Sources}

\author{Savvas Zannettou$^{\star}$, Tristan Caulfield$^{\dagger}$, Emiliano De Cristofaro$^\dagger$, Nicolas Kourtellis$^{\ddagger}$,\\ Ilias Leontiadis$^\ddagger$, Michael Sirivianos$^{\star}$, Gianluca Stringhini$^\dagger$, and Jeremy Blackburn$^+$}
\affiliation{\institution{$^{\star}$Cyprus University of Technology, $^\dagger$University College London, $^\ddagger$Telefonica Research,\\ ${^+}$University of Alabama at Birmingham}}
\affiliation{\institution{sa.zannettou@edu.cut.ac.cy, \{t.caulfield,e.decristofaro,g.stringhini\}@ucl.ac.uk, \{nicolas.kourtellis,ilias.leontiadis\}@telefonica.com, michael.sirivianos@cut.ac.cy, blackburn@uab.edu}}

\begin{abstract}
As the number and the diversity of news outlets on the Web grow, so does the opportunity for ``alternative'' sources of information to emerge.
Using large social networks like Twitter and Facebook, misleading, false, or agenda-driven information can quickly and seamlessly spread online, deceiving people or influencing their opinions.
Also, the increased engagement of tightly knit communities, such as Reddit and 4chan, further compounds the problem, as their users initiate and propagate alternative information, not only within their own communities, but also to different ones as well as various social media.
In fact, these platforms have become an important piece of the modern information ecosystem, which, thus far, has not been studied as a whole.

In this paper, we begin to fill this gap by studying mainstream and alternative news shared on Twitter, Reddit, and 4chan.
By analyzing millions of posts around several axes, we measure how mainstream and alternative news flows between these platforms.
Our results indicate that alt-right communities within 4chan and Reddit can have a surprising level of influence on Twitter, providing evidence that ``fringe'' communities often succeed in spreading alternative news to mainstream social networks and the greater Web.
\end{abstract}

\settopmatter{printfolios=true}
\maketitle

\section{Introduction}
Over the past few years, a number of high-profile conspiracy theories and false stories have originated and spread on the Web.
After the Boston Marathon bombings in 2013, a large number of tweets started to claim that the bombings were a ``false flag'' perpetrated by the United States government~\cite{starbird2014boston}.
Also, the GamerGate controversy started as a blogpost by a jaded ex-boyfriend that turned into a pseudo-political campaign of targeted online harassment~\cite{despoina2017websci}.
More recently, the Pizzagate conspiracy~\cite{pizzagate} -- a debunked theory connecting a restaurant and members of the US Democratic Party to a child sex ring -- led to a shooting in a Washinghton DC restaurant~\cite{nytimes_pizzagate_shooter}.
These stories were all propagated, in no small part, via the use of ``alternative'' news sites like Infowars and ``fringe'' Web communities like 4chan.
Overall, the barrier of entry for such alternative news sources has been greatly reduced by the Web and large social networks.
Due to the negligible cost of distributing information over social media, fringe sites can quickly gain traction with large audiences.
At the same time, the explosion of information sources also hinders the effective regulation of the sector, while further muddying the water when it comes to the evaluation of news information by readers.

While there are many plausible motives for the rise in alternative narratives~\cite{starbird2017examining}, ranging from libelous (e.g., to harm the image of a particular person or group), political (e.g,. to influence voters), profit (e.g., to make money from advertising), or trolling~\cite{allcott2017social}, the manner in which they proliferate throughout the Web is still unknown.
Although previous work has examined information cascades, rumors, and hoaxes~\cite{friggeri2014rumor, kumar2016disinformation,shao2016hoaxy}, to the
best of our knowledge, very little work provides a holistic view of the modern information ecosystem. 
This knowledge, however, is crucial for understanding the alternative news world and for designing appropriate detection/mitigation strategies.
Anecdotal evidence and press coverage suggest that alternative news dissemination might start on fringe sites, eventually reaching mainstream online social networks and news outlets~\cite{bbc_macron, bbc_4chan_pizzagate}.
Nevertheless, this phenomenon has not been measured and no thorough analysis has focused on how news moves from one online service to another.

In this paper, we address this gap by providing the first large-scale measurement of how mainstream and alternative news flows through multiple social media platforms.
We focus on the relationship between three fundamentally different %
social media platforms, Reddit, Twitter, and 4chan, which we choose because of: 
1) their fundamental differences as well as their generally accepted ``driving'' of substantial portions of the online world; 
2) anecdotal evidence that suggests that specific sub-communities within Reddit and 4chan act as generators~\cite{bbc_4chan_pizzagate} and incubators~\cite{guardian_reddit} of fake news stories; and 
3) the substantial impact they have in forming and manipulating peoples' opinions (and therefore actions), when they constantly disseminate false information~\cite{nytimes_pizzagate_shooter}.

\descr{Contributions.} %
First, we undertake a large-scale measurement and comparison of the occurrence of mainstream and alternative news sources across three social
media platforms (4chan, Reddit, and Twitter).
Then, we provide an understanding of the temporal dynamics of how URLs from news sites are posted on the different social networks.
Finally, we present a measurement of the \emph{influence} between the platforms that provides insight into how information spreads throughout
the greater Web.

Overall, our findings indicate that Twitter, Reddit, and 4chan are used quite extensively for the dissemination of both alternative and mainstream news.
Using a statistical model for influence -- namely, Hawkes processes -- we show that each of the platforms (and, in the case of Reddit, sub-communities) have varying degrees of influence on each other, and this influence differs with respect to mainstream and alternative news sources.
Naturally, our approach is not without limitations, which we discuss in details in Section~\ref{sec:conclusion}.

\descr{Paper Organization.} The rest of the paper is organized as follows.
The next section discusses the social networks and information sources studied in this paper.
Section~\ref{sec:general_characterization} presents a general characterization of each platform, while Section~\ref{sec:analysis} discusses
our temporal findings. Section~\ref{sec:influence_measurement} reports our measurements of the influence between the platforms. Finally, after reviewing related work in Section~\ref{sec:related_work}, the paper concludes in Section~\ref{sec:conclusion}.

\section{Background and Data Collection}
\label{sec:methodology}

In this section, we provide some background information on the three social media platforms we study, the selection of news sources, and details on the data we collect.

\subsection{Platforms and News Sources}

\descr{Twitter.} Twitter is a micro-blogging, directed social network where users  broadcast 140-character ``tweets'' to their followers.
Some of its features include the hashtag (a keyword preceded by \#), which makes it easier for users to find and weigh in on tweets around a theme, as well as retweeting, i.e., rebroadcasting a tweet.

\descr{Reddit.} The so-called ``front page of the Internet'' is a social news aggregator where users post URLs to content along with a
title, and other users can upvote or downvote the post.
Votes determine the ranking of the posts, i.e., the order in which they are displayed.
There is also a threaded comments section for users to discuss a post, and comments are also subject to the voting system. %
Although users can mark each other as friends, the community structure is not defined by the friendship relation. Rather, communities on Reddit are formed via the ``subreddit'' concept.
Users can create their own subreddits, choosing the topic as well as the moderation policy.
This has led to a plethora of communities, ranging from video games to news and politics,  pornography, and even meta-communities focusing on interactions people have in other subreddits.

\descr{4chan.} 4chan is a type of discussion forum known as an \emph{imageboard}: users create a new thread by making a post with a single image attached, and perhaps some text, in one of several boards (70 as of September 2017) for different topics of interest.
Other users can add posts to the thread, with or without an image, and quote or reply to posts.
Users are not required to provide a username to access or post to 4chan, and the default ``Anonymous'' is the preferred and overwhelmingly used identity. %
Another key characteristic of 4chan is ephemerality: there are a finite number of threads that can be active at a given time on a given board.
When a new thread is created, an old one is purged based on their ranking within the ``bump'' system~\cite{hine2016longitudinal}.
Although several boards have a temporary archive for purged posts, all threads are permanently deleted after 7 days.
4chan is known for its extremely lax moderation: while boards are divided into safe and not safe for work categories, volunteer ``janitors'' and paid employees  generally are not concerned with the language used or the tone of discussion, as long as the discussion falls within the general topic of the board.
Since 4chan's primary mode of operation is ``anonymous,'' it inherently lacks many of the ``social'' features of other social media platforms, and there is no concept of friends/followers.

In this work, we are primarily interested in the Politically Incorrect board, or \dspol, which focuses on the discussion of politics and world events, and has often been linked to the alt-right~\cite{medium}, exhibiting a high degree of racist and hate speech content~\cite{hine2016longitudinal}. We also use 4chan's Sports (\dssp), International (\dsint), and Science (\dssci) boards as a baseline.

\descr{News sites.} Our analysis uses a set of news websites that can confidently be labeled as either ``mainstream'' or ``alternative'' news.
More specifically, we create a list of 99 news sites including 45 mainstream and 54 alternative ones.\footnote{The complete list of the 99 sites is available at \url{https://drive.google.com/open?id=0ByP5a__khV0dM1ZSY3YxQWF2N2c}}
For the former, we select 45 from the Alexa top 100 news sites, leaving out those based on user-generated content, those serving specialized content
(e.g., finance news), as well as non-English sites.
For the latter, we use Wikipedia\footnote{\url{https://en.wikipedia.org/wiki/List\_of\_fake\_news\_websites}} and
FakeNewsWatch.\footnote{\url{http://fakenewswatch.com/}}
We also add two state-sponsored alternative news domains: \url{sputniknews.com} and \url{rt.com}, as they have recently attracted public attention due to their posting of controversial, and seemingly agenda-pushing stories~\cite{sputnik_spotlight2}.

\begin{table}[t]
\centering
\small
\begin{tabular}{lrrr}
\hline
\textbf{Platform}         & \textbf{Total Posts} & \textbf{\% Alt.} & \textbf{\% Main.} \\ \hline
Twitter                   & 587M   & 0.022\%     & 0.070\%   \\
Reddit (posts + comments) & 332M   & 0.023\%     & 0.181\%   \\
4chan                     & 42M   & 0.050\%     & 0.197\%   \\ \hline
\end{tabular}
\caption{Total number of posts crawled and percentage of posts that contain URLs to our list of alternative and mainstream news sites.}
\label{tbl:post_percentage}
\end{table}

\begin{table}[t]
\centering
\resizebox{\columnwidth}{!}{
\begin{tabular}{lrrr}
\hline
\textbf{Platform}    & \hspace*{-0.6cm} \textbf{Posts/Comments} & \textbf{Alt. URLs} & \textbf{Main. URLs} \\ \hline
Twitter              & 486,700                       & 42,550                    & 236,480                  \\
Reddit (six selected subreddits)\hspace*{-3cm}      & 620,530                       & 40,046                    & 301,840                  \\
Reddit (all other subreddits)\hspace*{-3cm}  & 1,228,105                     & 24,027                    & 726,948                  \\
4chan (\dspol)        & 90,537                        & 8,963                     & 40,164                   \\
4chan (\dsint, \dssci, \dssp) & 7,131                         & 615                       & 5,513                    \\ \hline
\end{tabular}
}
\caption{Number of posts/comments that contain a URL to one of our information sources, as well as the number of unique URLs linking to alternative and mainstream news sites in our list.}
\label{tbl:datasets}
\end{table}

\begin{table}[t]
\centering
\small
\setlength{\tabcolsep}{3.5pt}
\begin{tabular}{@{}lrrrrrr@{}}
\hline
   & {\bf Tweets} & {\bf Retrieved (\%)}  &  {\bf Avg. Retweets} & {\bf Avg.~Likes}   \\ 
   \hline
\textbf{Alternative} & 110,629  & 92,104 (83.2\%)  & 341 $\pm$ 1,228           & 0.82 $\pm$ 15.6                           \\
\textbf{Mainstream}  & 376,071  & 329,950 (87.7\%) & 404 $\pm$ 2,146          & 0.96 $\pm$ 55.6                         \\ \hline
\end{tabular}
\caption{Statistics of alternative and mainstream news URLs in the tweets in our dataset.}
\label{table-tweets-stats}
\end{table}

\subsection{Datasets}\label{sec:data_collection}
We gather information from posts, threads, and comments on Twitter, Reddit, and 4chan that contain URLs from the 99 news sites. %
With a few gaps (see below), our datasets cover activity on the three platforms between June 30, 2016 and February 28, 2017.
Table~\ref{tbl:post_percentage} shows the total number of posts/comments crawled and the percentage of posts that contains links to URLs from the aforementioned news domains.
We observe that mainstream news URLs are present in a greater percentage of posts on 4chan and Reddit than on Twitter,
while alternative ones are about twice as likely to appear in posts on 4chan than on Twitter or Reddit.
Table~\ref{tbl:datasets} provides a summary of our datasets, which we present in more detail below.
Note that we break Reddit and 4chan datasets into two different instances, as further discussed in Section~\ref{sec:general_characterization}.

\descr{Twitter.} We collect the 1\% of all publicly available tweets with URLs from the aforementioned news domains between June 30, 2016 and February 28, 2017 using the Twitter Streaming
API.\footnote{\url{https://dev.twitter.com/streaming/overview}}
In total, we gather 487k tweets containing 279k unique URLs pointing to mainstream or alternative news sites. 
Since tweets are retrieved at the time they are posted, we do not get information such as the number of times they are re-tweeted or liked.
Therefore, between March and May 2017, we re-crawled each tweet to retrieve this data.
Basic statistics are summarized in Table~\ref{table-tweets-stats}.
Due to a failure in our collection infrastructure, we have some gaps in the Twitter dataset, specifically between Oct 28--Nov 2 and Nov 5--16, 2016, as well as Nov 22, 2016 -- Jan 13, 2017, and Feb 24--28, 2017.

\begin{table}[]
\centering
\small
\begin{tabular}{rcrr}
\hline
\textbf{Subreddit (Alt.)} & \textbf{(\%)}       & \textbf{Subreddit (Main.)} & \textbf{ (\%)} \\ \hline
The\_Donald                  & \multicolumn{1}{r|}{35.37 \%} & politics                    & 12.9 \%                   \\
politics                   & \multicolumn{1}{r|}{8.21 \%}  & worldnews                   & 6.24 \%                   \\
news                       & \multicolumn{1}{r|}{3.85 \%}  & The\_Donald                 & 4.53 \%                   \\
conspiracy                 & \multicolumn{1}{r|}{3.84 \%}  & news                        & 4.23 \%                   \\
Uncensored                 & \multicolumn{1}{r|}{2.66 \%}  & TheColorIsBlue              & 3.06 \%                   \\
Health                     & \multicolumn{1}{r|}{2.10 \%}  & TheColorIsRed               & 2.48 \%                   \\
PoliticsAll                & \multicolumn{1}{r|}{1.54 \%}  & willis7737\_news            & 2.27 \%                   \\
Conservative               & \multicolumn{1}{r|}{1.45 \%}  & news\_etc                   & 1.94 \%                   \\
worldnews                  & \multicolumn{1}{r|}{1.41 \%}  & AskReddit                   & 1.37 \%                   \\
WhiteRights                & \multicolumn{1}{r|}{1.21 \%}  & canada                      & 1.31 \%                   \\
KotakuInAction             & \multicolumn{1}{r|}{1.04 \%}  & EnoughTrumpSpam             & 1.20 \%                   \\
HillaryForPrison           & \multicolumn{1}{r|}{0.94 \%}  & NoFilterNews                & 1.16 \%                   \\
TheOnion                   & \multicolumn{1}{r|}{0.94 \%}  & BreakingNews24hr            & 1.07 \%                   \\
AskTrumpSupporters         & \multicolumn{1}{r|}{0.84 \%}  & conspiracy                  & 0.89 \%                   \\
POLITIC                    & \multicolumn{1}{r|}{0.81 \%}  & todayilearned               & 0.83 \%                   \\
rss\_theonion              & \multicolumn{1}{r|}{0.67 \%}  & thenewsrightnow             & 0.78 \%                   \\
the\_Europe                & \multicolumn{1}{r|}{0.67 \%}  & europe                      & 0.77 \%                   \\
new\_right                 & \multicolumn{1}{r|}{0.60 \%}   & ReddLineNews                & 0.75 \%                   \\
AskReddit                  & \multicolumn{1}{r|}{0.59 \%}  & hillaryclinton              & 0.73 \%                  \\
AnythingGoesNews           & \multicolumn{1}{r|}{0.51 \%}  & nottheonion                 & 0.73 \%                  \\ \hline
\end{tabular}
\caption{Top 20 subreddits w.r.t. mainstream and alternative news URLs occurrence, and their percentage, in Reddit.}

\label{top_subreddits}
\end{table}
\descr{Reddit.} We obtain all posts and comments on Reddit between June 30, 2016 and February 28, 2017, using data made available on Pushshift.\footnote{\url{http://files.pushshift.io/}}
We collect approximately 42M posts, 390M comments, and 300k subreddits.
Once again, we filter posts and comments that contain URLs from one of the 99 news sites, which yields a dataset of 1.8M posts/comments and approximately 1.1M URLs.

\descr{4chan.} For 4chan, we use all threads and posts made on the Politically Incorrect (\dspol)  board, as well as \dssp (Sports), \dsint (International), and \dssci (Science) boards for comparison,
using the same methodology as~\cite{hine2016longitudinal}.
We opt to select both not safe for work boards (i.e., \dspol) and safe for work boards (i.e., \dssp, \dsint, and \dssci) to observe how these compare to each other with respect to the dissemination of news.
The resulting dataset includes 97k posts and replies, including 56k alternative and mainstream news URLs, between June 30, 2016 and February 28, 2017. We have some small gaps due to our crawler failing, specifically, Oct 15--16 and Dec 16--25, 2016 as well as Jan 10--13, 2017.

\section{General Characterization}
\label{sec:general_characterization}
In this section, we present a general characterization of the mainstream and alternative news URLs found on the three platforms.
\descr{Reddit.} We start by identifying news and politics communities.
In Table~\ref{top_subreddits}, we report the top 20 subreddits with the most URLs, along with their percentage.
Note that we omit automated ones (e.g., /r/AutoNewspaper/) where news articles are posted without user intervention.
Many of the subreddits are indeed related to news and politics -- e.g.,
`The\_Donald' is mostly a community of Donald Trump supporters, while `worldnews' is focused around globally relevant events.
We also find the presence of the `conspiracy' subreddit, which has been involved in disinformation campaigns including Pizzagate, %
as well as  `AskReddit,' where both mainstream and alternative news sources are used
to answer questions submitted by users. Although the latter is intended for open-ended questions that spark discussion, it is evident that commenters often try to push their agenda even in non-political threads.
In the end, based on their propensity to include news URLs of both types, we single out the follow top six subreddits for further exploration:
The\_Donald, politics, conspiracy, news, worldnews, and AskReddit.

\begin{table}[t]
\centering
\small
\begin{tabular}{@{}rcrc@{}}
\hline
\textbf{Domain (Alt.)}   & \textbf{(\%)}    & \textbf{Domain (Main.)} & \textbf{(\%)} \\ \hline
breitbart.com            & \multicolumn{1}{r|}{55.58 \%} & nytimes.com             & 14.07 \%                    \\
rt.com                   & \multicolumn{1}{r|}{19.18 \%} & cnn.com                 & 11.23 \%                    \\
infowars.com             & \multicolumn{1}{r|}{8.99 \%}  & theguardian.com         & 8.86 \%                     \\
sputniknews.com          & \multicolumn{1}{r|}{3.95 \%}  & reuters.com             & 6.67 \%                     \\
beforeitsnews.com        & \multicolumn{1}{r|}{2.34 \%}  & huffingtonpost.com      & 5.67 \%                     \\
lifezette.com            & \multicolumn{1}{r|}{2.28 \%}  & thehill.com             & 5.15 \%                     \\
naturalnews.com          & \multicolumn{1}{r|}{1.54 \%}  & foxnews.com             & 4.89 \%                     \\
activistpost.com         & \multicolumn{1}{r|}{1.45 \%}  & bbc.com                 & 4.76 \%                     \\
veteranstoday.com        & \multicolumn{1}{r|}{1.11 \%}  & abcnews.go.com          & 2.94 \%                     \\
redflagnews.com          & \multicolumn{1}{r|}{0.63 \%}  & usatoday.com            & 2.87 \%                     \\
prntly.com               & \multicolumn{1}{r|}{0.49 \%}  & nbcnews.com             & 2.86 \%                     \\
dccclothesline.com       & \multicolumn{1}{r|}{0.4 \%}   & time.com                & 2.57 \%                     \\
worldnewsdailyreport.com & \multicolumn{1}{r|}{0.36 \%}  & washinghtontimes.com    & 2.52 \%                     \\
therealstrategy.com      & \multicolumn{1}{r|}{0.3 \%}   & bloomberg.com           & 2.5 \%                      \\
disclose.tv              & \multicolumn{1}{r|}{0.23 \%}  & wsj.com                 & 2.31 \%                     \\
clickhole.com            & \multicolumn{1}{r|}{0.2 \%}   & cbsnews.com             & 2.26 \%                     \\
libertywritersnews.com   & \multicolumn{1}{r|}{0.2 \%}   & thedailybeast.com       & 2.05 \%                     \\
worldtruth.tv            & \multicolumn{1}{r|}{0.14 \%}  & forbes.com              & 1.87 \%                     \\
thelastlineofdefence.org & \multicolumn{1}{r|}{0.07 \%}  & nypost.com              & 1.85 \%                     \\
nodisinfo.com            & \multicolumn{1}{r|}{0.05 \%}  & cncb.com                & 1.54 \%                     \\ \hline
\end{tabular}
\caption{Top 20 mainstream and alternative news sites and their percentage (six selected subreddits).}
\label{tbl:reddit_top_domains}
\end{table}

In order to get a better view of the popularity of news sites on the six subreddits, we study the occurrence of each news outlet.
Specifically, we find 76k URLs (40k unique) from alternative news
and 600k (301k unique) from mainstream news domains.
Table~\ref{tbl:reddit_top_domains} reports the top 20 mainstream/alternative news sites and their percentage in the six subreddits.
The top 20 domains for mainstream news account for 89\% of all mainstream news URLs in our data, while for alternative domains the percentage is 99\%.
Known alt-right news outlets, such as \url{breitbart.com} and \url{infowars.com}, are predominantly present, as well as state-sponsored
alternative domains like \url{sputniknews.com} and \url{rt.com}, which have recently been in the spotlight for disseminating false information
and propaganda~\cite{sputnik_spotlight2}.  %
The fact that many such URLs appear in our dataset may indeed be an indication that the six subreddits significantly contribute to the dissemination of controversial stories.

\begin{table}[t]
\centering
\small
\setlength{\tabcolsep}{2pt}
\begin{tabular}{rrrc}
\hline
\textbf{Domain (Alt.)}   & \textbf{(\%)}    & \textbf{Domain (Main.)} & \textbf{(\%)} \\ \hline
breitbart.com            & \multicolumn{1}{r|}{46.04 \%} & theguardian.com         & 19.04 \%                    \\
rt.com                   & \multicolumn{1}{r|}{17.56 \%} & nytimes.com             & 10.07 \%                    \\
infowars.com             & \multicolumn{1}{r|}{17.25 \%} & bbc.com                 & 8.99 \%                     \\
therealstrategy.com      & \multicolumn{1}{r|}{5.63 \%}  & forbes.com              & 6.24 \%                     \\
sputniknews.com          & \multicolumn{1}{r|}{4.11 \%}  & thehill.com             & 4.95 \%                     \\
beforeitsnews.com        & \multicolumn{1}{r|}{2.26 \%}  & cbc.ca                  & 4.82 \%                     \\
redflagnews.com          & \multicolumn{1}{r|}{2.04 \%}  & foxnews.com             & 4.79 \%                     \\
dccclothesline.com       & \multicolumn{1}{r|}{1.37 \%}  & wsj.com                 & 4.04 \%                     \\
naturalnews.com          & \multicolumn{1}{r|}{1.29 \%}  & bloomberg.com           & 3.48 \%                     \\
clickhole.com            & \multicolumn{1}{r|}{0.53 \%}  & reuters.com             & 2.85 \%                     \\
activistpost.com         & \multicolumn{1}{r|}{0.41 \%}  & usatoday.com            & 2.02 \%                     \\
disclose.tv              & \multicolumn{1}{r|}{0.39 \%}  & thedailybeast.com       & 2.02 \%                     \\
prntly.com               & \multicolumn{1}{r|}{0.26 \%}  & nbcnews.com             & 1.96 \%                     \\
worldtruth.tv            & \multicolumn{1}{r|}{0.25 \%}  & nypost.com              & 1.95 \%                     \\
libertywritersnews.com   & \multicolumn{1}{r|}{0.15 \%}  & cbsnews.com             & 1.89 \%                     \\
worldnewsdailyreport.com & \multicolumn{1}{r|}{0.06 \%}  & abcnews.go.com          & 1.78 \%                     \\
mediamass.net            & \multicolumn{1}{r|}{0.04 \%}  & time.com                & 1.71 \%                     \\
newsbiscuit.com          & \multicolumn{1}{r|}{0.03 \%}  & cnbc.com                & 1.40 \%                     \\
react365.com             & \multicolumn{1}{r|}{0.02 \%}  & washingtontimes.com     & 1.34 \%                     \\
the-daily.buzz           & \multicolumn{1}{r|}{0.02 \%}  & washingtonexaminer.com  & 1.33 \%                     \\ \hline
\end{tabular}
\caption{Top 20 mainstream and alternative news sites, and their percentage, in the Twitter dataset.}
\label{twitter_top_domains}
\end{table}

\descr{Twitter.} In our Twitter dataset, we find 129k (42k unique) URLs of alternative news domains and 413k (236k unique) URLs of
mainstream ones. Recall that we re-crawl
tweets to get the number of retweets and likes, and a small percentage of them are no longer available as they were either deleted
or the associated account was suspended.
This percentage is slightly higher for tweets with URLs from alternative news, possibly due to the fact that some users tend to remove
controversial content when a particular false story is debunked~\cite{friggeri2014rumor}.
Also, alternative and mainstream news tend to get a significant number of retweets, at about the same rate (on average, 404 and 341 retweets
per tweet, respectively). A similar pattern is observed for likes (see Table~\ref{table-tweets-stats}).

In Table~\ref{twitter_top_domains}, we report the top 20 mainstream and alternative news domains, and their percentage, in our Twitter dataset.
These cover, respectively, 86\% and 99\% of all URLs.
Similar to Reddit, there are many popular alt-right and state-sponsored news outlets.

\begin{table}[t]
\centering
\small
\setlength{\tabcolsep}{2pt}
\begin{tabular}{rrrr}
\hline
\textbf{Domain (Alt.)}   & \textbf{(\%)}    & \textbf{Domain (Main.)} & \textbf{(\%)} \\ \hline
breitbart.com            & \multicolumn{1}{r|}{53.00 \%} & theguardian.com         & 14.10 \%                    \\
rt.com                   & \multicolumn{1}{r|}{28.22 \%} & nytimes.com             & 10.07 \%                    \\
infowars.com             & \multicolumn{1}{r|}{9.12 \%}  & cnn.com                 & 9.90 \%                     \\
sputniknews.com          & \multicolumn{1}{r|}{3.36 \%}  & bbc.com                 & 5.45 \%                     \\
veteranstoday.com        & \multicolumn{1}{r|}{1.07 \%}  & foxnews.com             & 5.35 \%                     \\
beforeitsnews.com        & \multicolumn{1}{r|}{0.91 \%}  & reuters.com             & 5.10 \%                     \\
lifezette.com            & \multicolumn{1}{r|}{0.86 \%}  & time.com                & 3.42 \%                     \\
naturalnews.com          & \multicolumn{1}{r|}{0.61 \%}  & abcnews.go.com          & 3.40 \%                     \\
worldnewsdailyreport.com & \multicolumn{1}{r|}{0.46 \%}  & huffingtonpost.com      & 3.29 \%                     \\
prntly.com               & \multicolumn{1}{r|}{0.41 \%}  & thehill.com             & 3.04 \%                     \\
activistpost.com         & \multicolumn{1}{r|}{0.38 \%}  & wsj.com                 & 2.82 \%                     \\
dccclothesline.com       & \multicolumn{1}{r|}{0.29 \%}  & washinghtontimes.com    & 2.77 \%                     \\
redflagnews.com          & \multicolumn{1}{r|}{0.20 \%}  & bloomberg.com           & 2.75 \%                     \\
libertywritersnews.com   & \multicolumn{1}{r|}{0.16 \%}  & cbc.ca                  & 2.66 \%                     \\
therealstrategy.com      & \multicolumn{1}{r|}{0.16 \%}  & nypost.com              & 2.65 \%                     \\
clickhole.com            & \multicolumn{1}{r|}{0.11 \%}  & cbsnews.com             & 2.44 \%                     \\
disclose.tv              & \multicolumn{1}{r|}{0.10 \%}  & nbcnews.com             & 2.32 \%                     \\
now8news.com             & \multicolumn{1}{r|}{0.06 \%}  & usatoday.com            & 2.25 \%                     \\
firebrandleft.com        & \multicolumn{1}{r|}{0.05 \%}  & cnbc.com                & 2.13 \%                     \\
nodisinfo.com            & \multicolumn{1}{r|}{0.05 \%}  & forbes.com              & 1.68 \%                     \\ \hline
\end{tabular}
\caption{Top 20 mainstream and alternative news sites, and their percentage, in the /pol/ dataset.}
\label{tbl:4chan_top_domains}
\end{table}

\descr{4chan.}
In our \dspol dataset, we find 21k (9k unique) URLs to alternative news outlets and 82k (40k unique) to mainstream news.
Table~\ref{tbl:4chan_top_domains} reports the percentage of URLs of the top 20 domains for each type of news.
These cover 87\% and 99\% of mainstream and alternative news URLs, respectively.
Again, we observe that, by far, the most popular alternative news domains are \url{breitbart.com}, \url{rt.com},
\url{infowars.com}, and \url{sputniknews.com}.
For the mainstream news, we observe that \url{theguardian.com} is the most frequently posted, followed by \url{nytimes.com}, \url{cnn.com}, and \url{bbc.com}.
We also obtained similar statistics for domain popularity in the other boards of 4chan, but we omit them for brevity.

\begin{figure}[t]
\center
\subfigure[]{\includegraphics[width=0.37\textwidth]{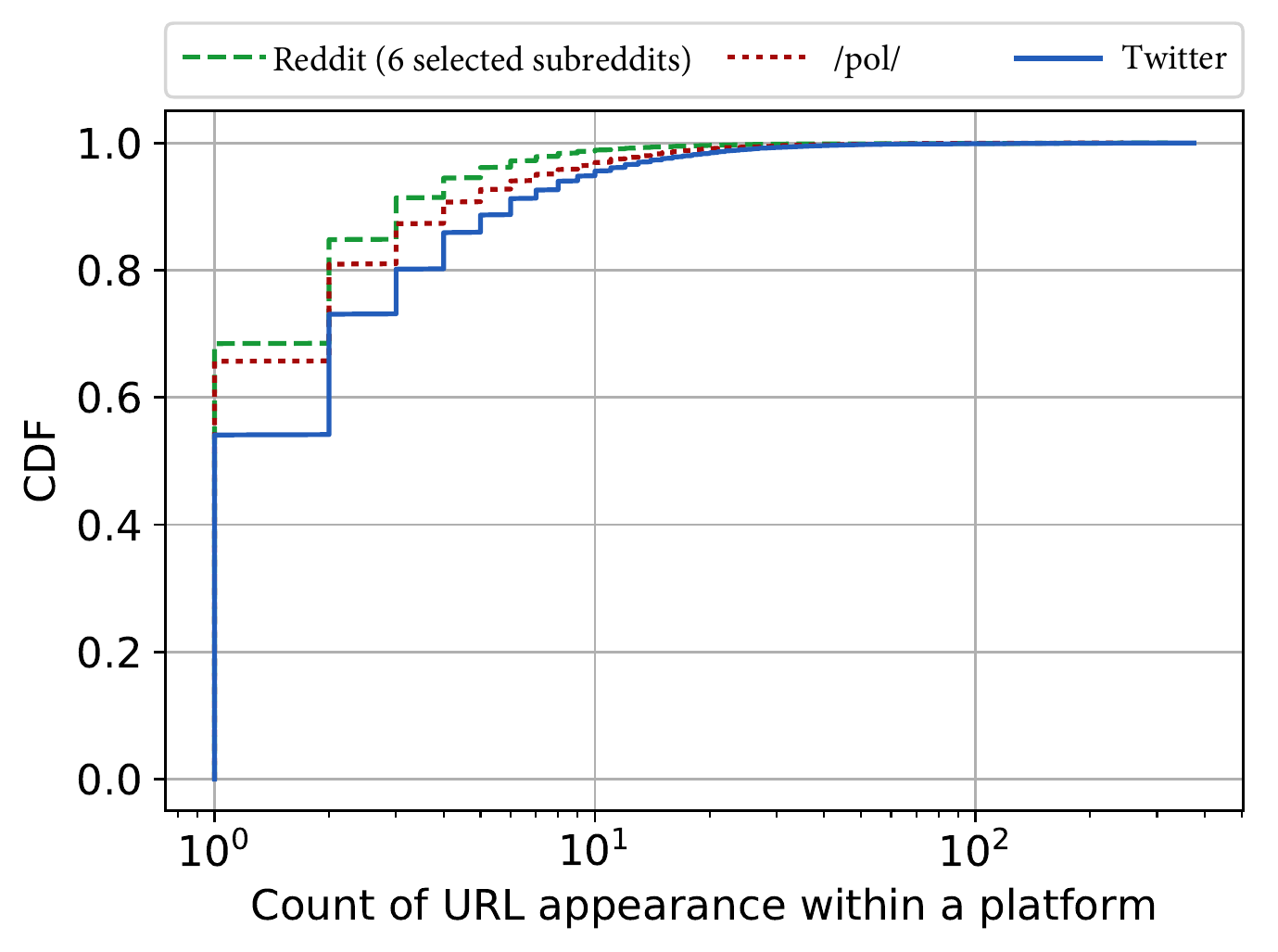}\label{url-occurence-fake}}\\[-1ex]
\subfigure[]{\includegraphics[width=0.37\textwidth]{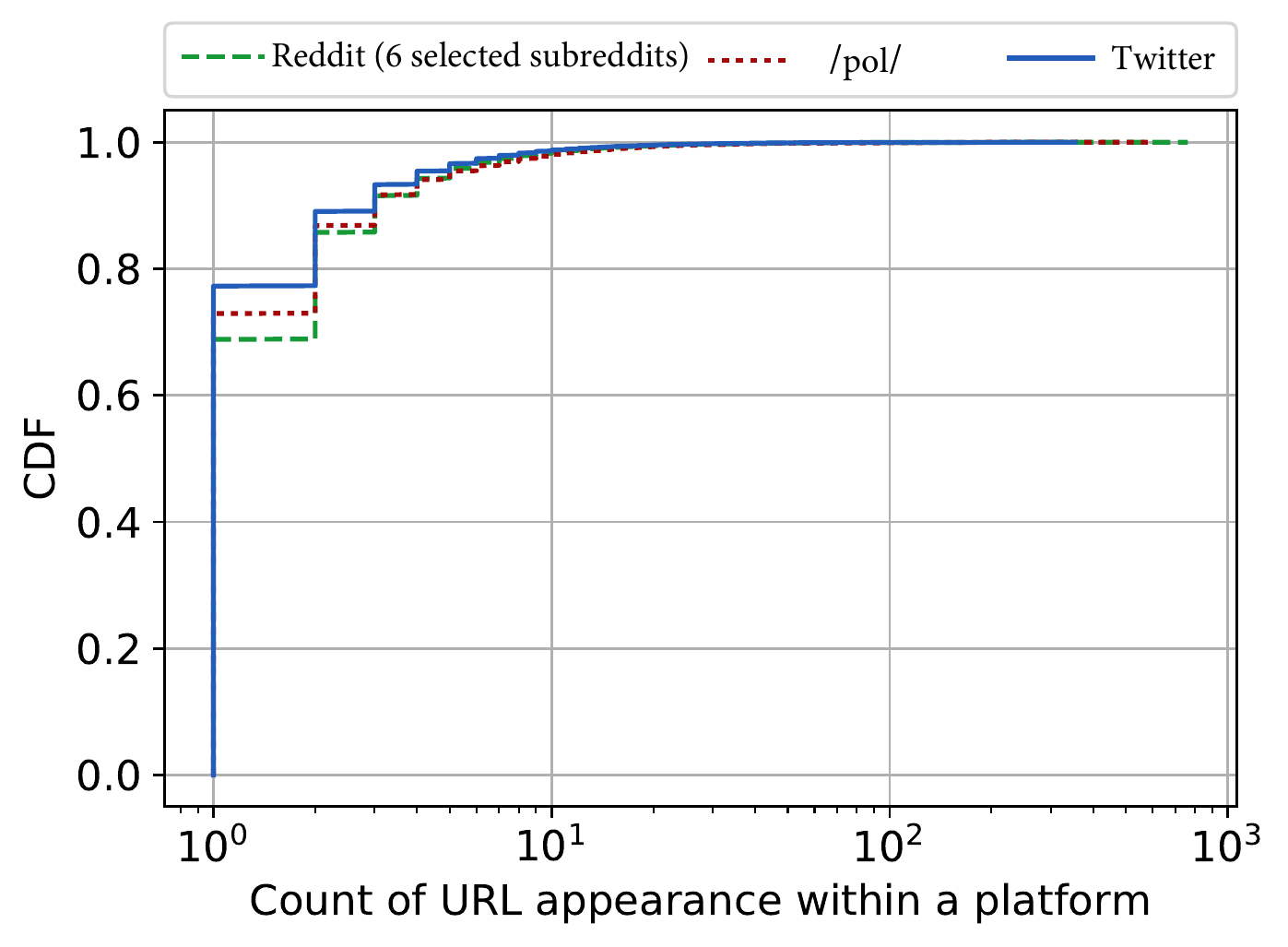}\label{url-occurence-real}}
\caption{CDF of URL appearance counts within a platform: (a) alternative news and (b) mainstream news. }
\label{cdf_url_occurence}
\end{figure}

\begin{figure*}[t]
\center
\subfigure[]{\includegraphics[width=0.44\textwidth]{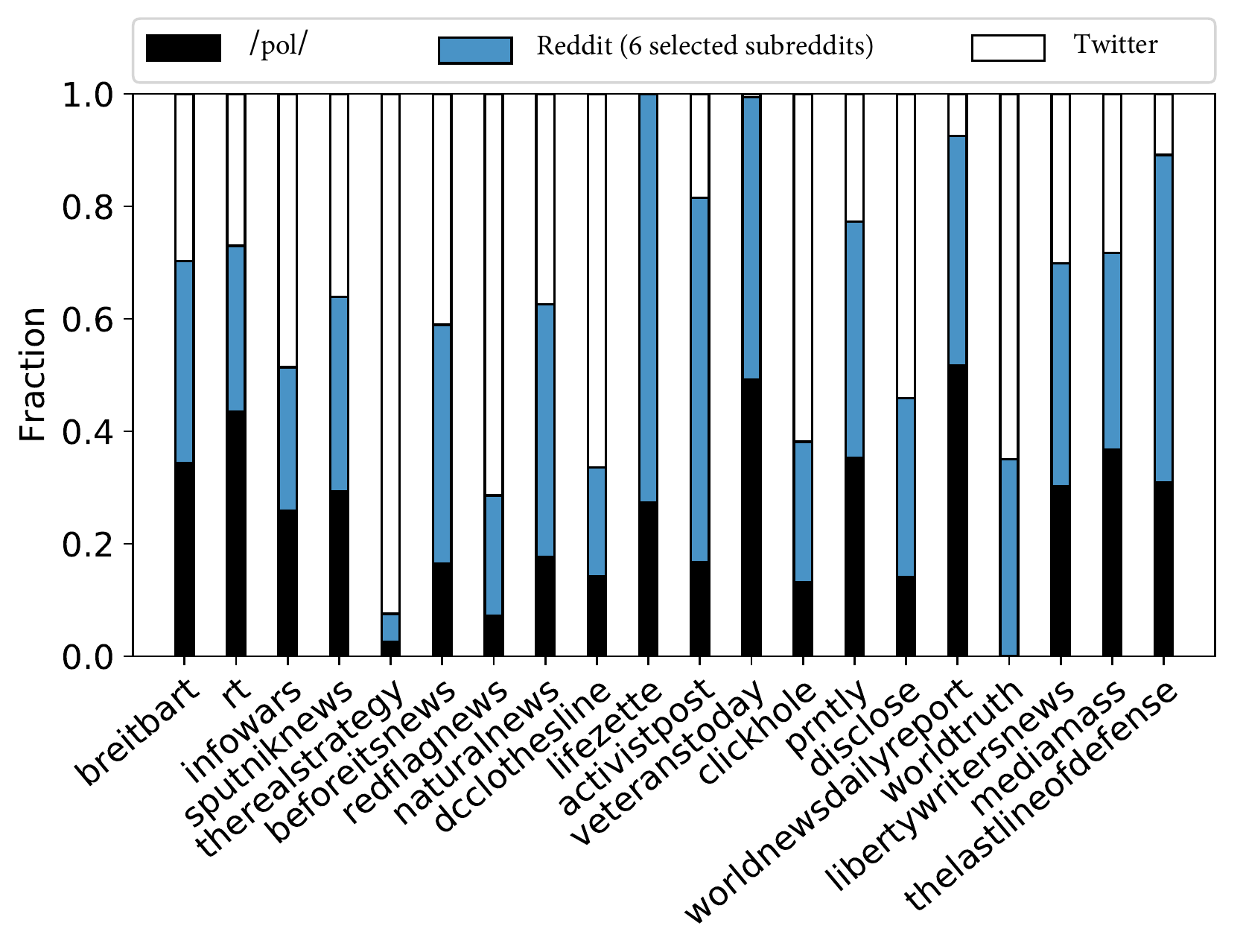}\label{domain_ratio_platforms-fake}}
~
\subfigure[]{\includegraphics[width=0.44\textwidth]{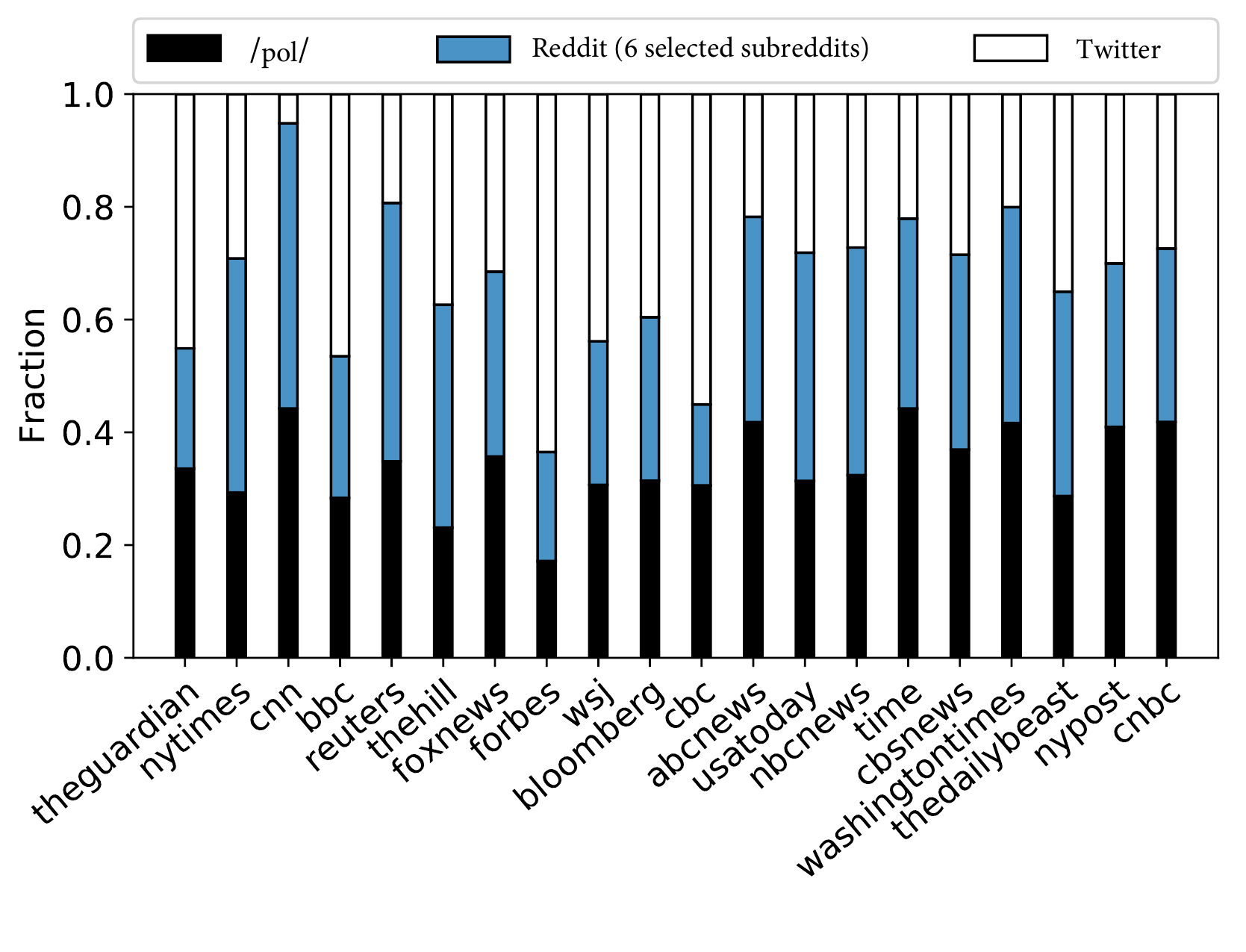}\label{domain_ratio_platforms-real}}
\caption{Top 20 domains and each platform's fraction for (a) alternative and (b) mainstream news.}
\label{domain_ratio_platforms}
\end{figure*}

To get a better view of the platforms' URL posting behavior, Figure~\ref{cdf_url_occurence} plots the CDF of URL appearances (i.e., how many times a specific URL appears) within a particular platform.
We observe that a substantial portion of the URLs appear only once for both alternative and mainstream news, and that, on Twitter, alternative news tends to appear more times than  mainstream news. For \dspol and the six subreddits, we observe a similar behavior for both mainstream and alternative news.

\begin{figure}[t]
\center
\subfigure[ ]{\includegraphics[width=0.37\textwidth]{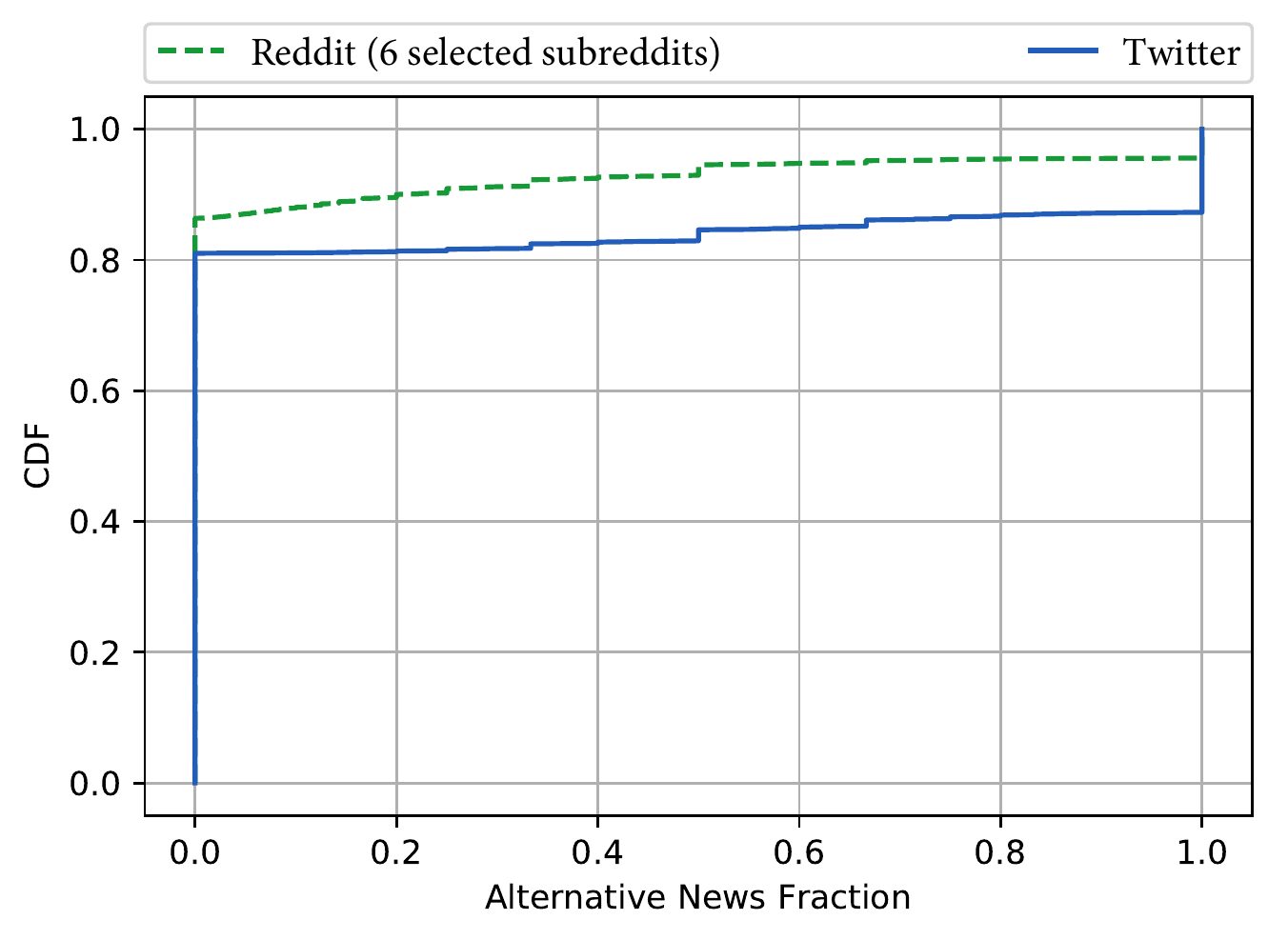}\label{fake_real_ratio_cdf_all}}\\[-1ex]
\subfigure[]{\includegraphics[width=0.37\textwidth]{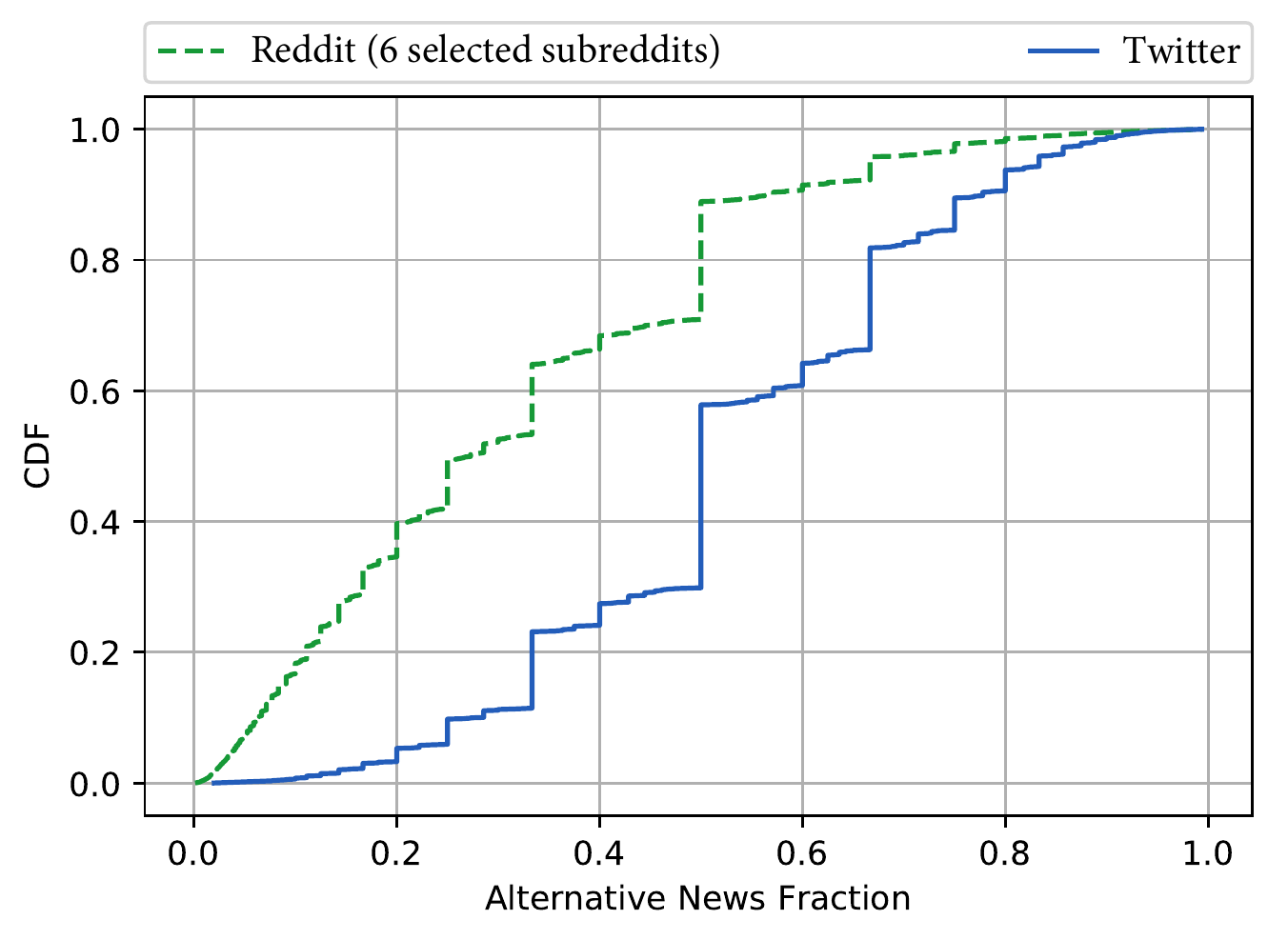}\label{fake_real_ratio_cdf_common}}
\caption{CDF of the fraction of URLs from alternative news and overall news URLs for (a) all users in our Twitter and Reddit datasets, and (b) users that shared URLs from both mainstream and alternative news.}
\label{fake_real_ratio_cdf}
\end{figure}

\begin{figure*}[t]
\center
\subfigure[]{\includegraphics[width=0.337\textwidth]{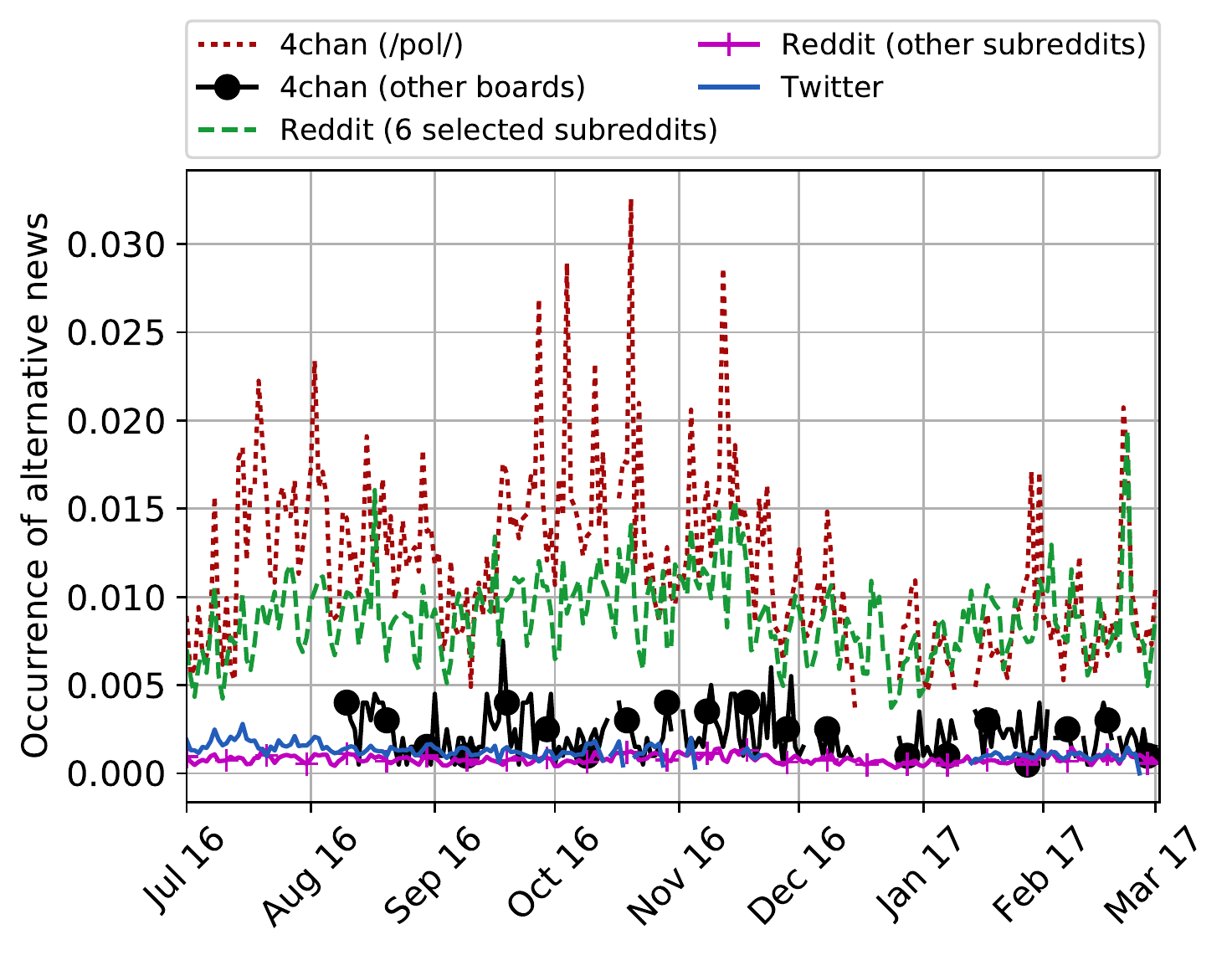}\label{aggregate_month_counts-fake}}
\hspace*{-0.3cm}
\subfigure[]{\includegraphics[width=0.337\textwidth]{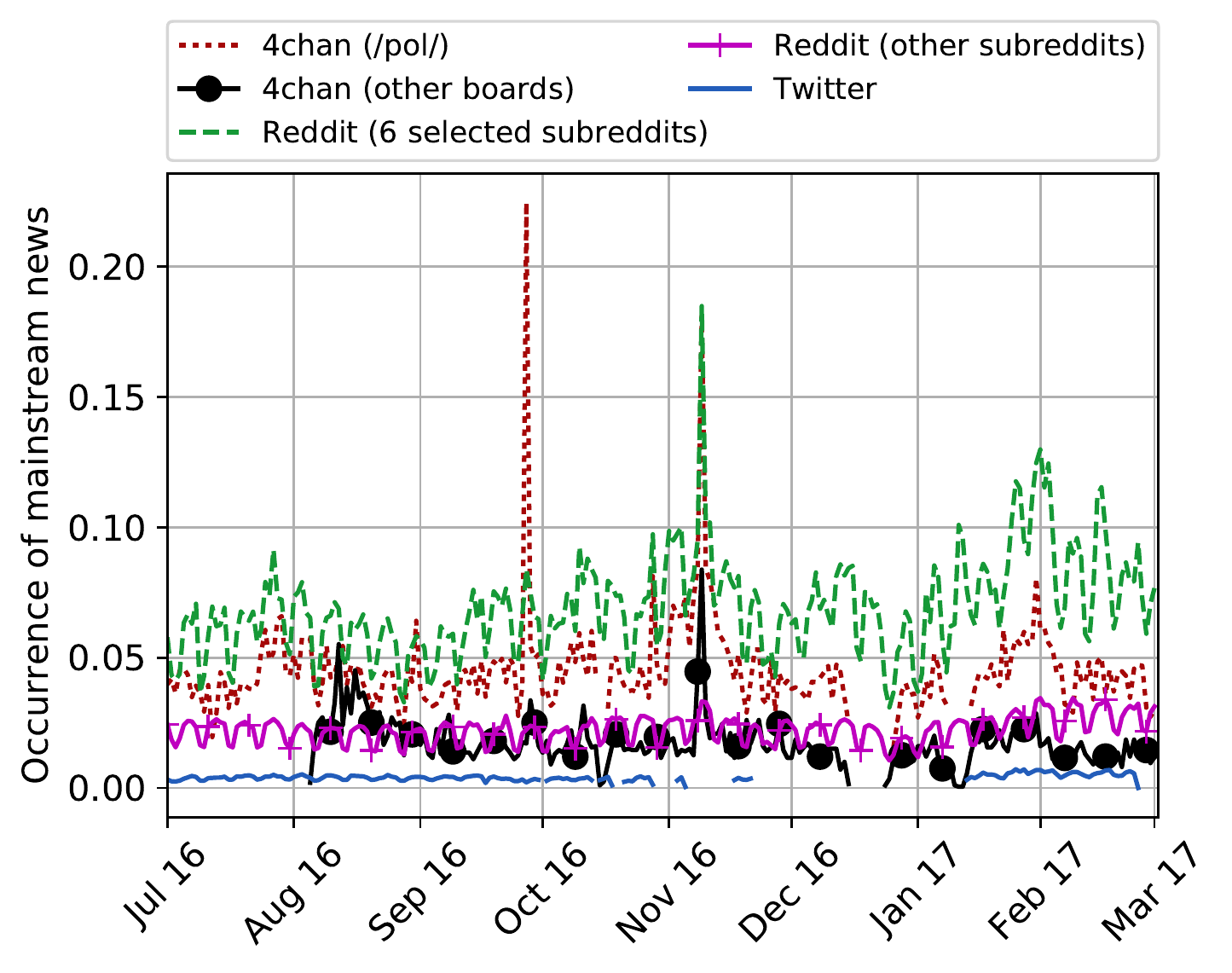}\label{aggregate_month_counts-real}}
\hspace*{-0.3cm}
\subfigure[]{\includegraphics[width=0.337\textwidth]{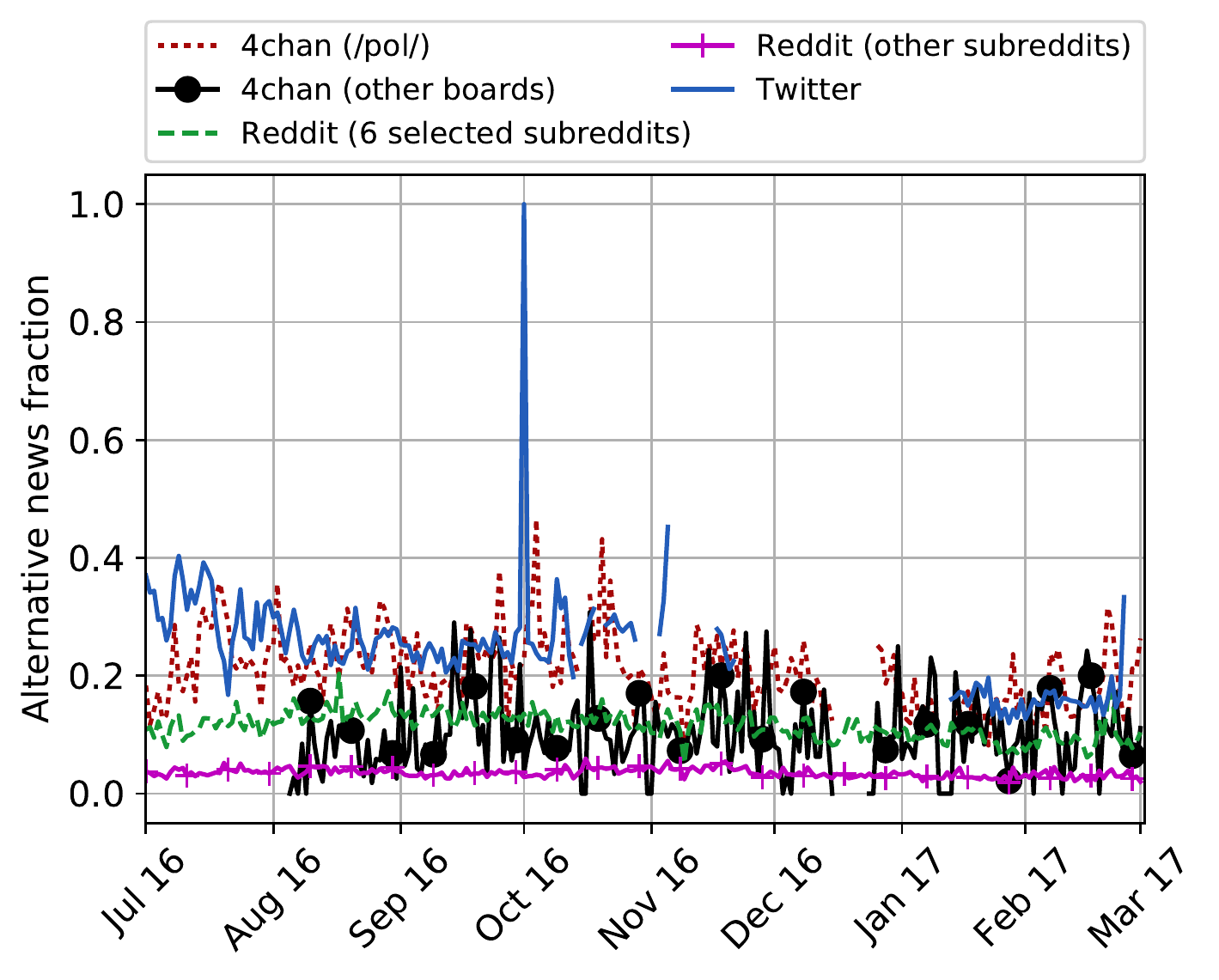}\label{aggregate_month_counts-ratio}}
\caption{Normalized daily occurrence of URLs for (a) alternative news, (b) mainstream news, and (c) fraction of alternative news over all news.}
\label{aggregate_month_counts}
\end{figure*}
Next, in Figure~\ref{domain_ratio_platforms}, we compare how popular domains, in both categories, appear on the three platforms
(i.e., Twitter, the six subreddits, and \dspol).
We find that the top 4 alternative domains -- \url{breitbart.com}, \url{rt.com}, \url{infowars.com}, \url{sputniknews.com} -- influence the three platforms more or less in the same way.
However, some outlets appear predominantly in some platforms but not in others; e.g., \url{therealstrategy.com} is popular only on Twitter,
while \url{lifezette.com} and \url{veteranstoday.com} are popular on the 6 subreddits and \dspol, but not on Twitter.

We believe the primary reason for this has to do with %
Twitter bots.
We cannot exclude with certainty that bots do not exist on 4chan, while bots are actually acceptable on Reddit (as long as they follow the rules of Reddit's API~\cite{reddit_api_rules}), however, they are certainly more prevalent on Twitter.
Thus, if a particular domain is popular on Twitter because of the influence of bots, then it might not be popular on Reddit and 4chan.
We have also considered ways to factor out posting behavior from bots, especially for Twitter, such as the one proposed in~\cite{davis2016botornot}. However, we have not removed this activity due to: 
1) posting behavior from bots can affect real users' posting behavior, hence this activity is part of the overall news dissemination ecosystem and needs to be accounted for; 
and 2) the satisfactory performance of such approaches is yet to be proven.

We also measure the fraction of news URLs that are alternative, {\em per user}, in Figure~\ref{fake_real_ratio_cdf}.
We report this fraction only for Reddit and Twitter users, since on 4chan posts are anonymous.
We find that 80\% of the users of both platforms share only URLs from mainstream news,
while, 13\% of Twitter users -- which are likely bots~\cite{varol2017online} -- exclusively post URLs to alternative news.
We observe from Figure~\ref{fake_real_ratio_cdf_common}, which shows the ratio for users sharing URLs from both categories, that there is a wide distribution, especially on the six selected subreddits, between people that rarely share alternative news (fraction close to 0) and those who share them almost all the time (fraction close to 1).
Moreover, we find that Twitter users share more alternative news: just 5\% of these users have a fraction below 0.2, %
which might be also attributed to the presence of bots.

\descr{Take-Aways.} In summary, our general characterization yields the following findings:
1) Specific sub-communities within Reddit drive the dissemination of both alternative and mainstream news (Table~\ref{top_subreddits});
2) news domain popularity is similar for both alternative and mainstream domains in the three platforms with some exceptions, such as \url{lifezette.com} and \url{veteranstoday.com} (Tables~\ref{tbl:reddit_top_domains},~\ref{twitter_top_domains},~\ref{tbl:4chan_top_domains} and Figure~\ref{domain_ratio_platforms});
3) Twitter users are more aggressively promoting alternative news, compared to mainstream news (Figure~\ref{cdf_url_occurence}); and
4) Twitter users also have a greater alternative to mainstream news ratio when compared to users that post in the six selected subreddits (Figure~\ref{fake_real_ratio_cdf}).

\section{Temporal Dynamics}
\label{sec:analysis}
In this section, we present the results of a cross-platform temporal analysis of the way news are posted on Twitter, Reddit, and 4chan.

\subsection{URL Occurrence}
In Figure~\ref{aggregate_month_counts}, we measure the daily occurrence of news URLs over the three platforms normalized by the average daily number of URLs shared in each community.\footnote{Gaps in the plot correspond to gaps in our dataset due to crawler failure.}
We find that \dspol and the six selected subreddits exhibit a much higher percentage of occurrences of alternative news compared to the other communities (Figure~\ref{aggregate_month_counts-fake}), whereas, for mainstream news, the sharing behavior is more similar across platforms (Figure~\ref{aggregate_month_counts-real}).
There are also some interesting spikes, likely related to the 2016 US elections, on the date of the first presidential debate and election day itself.
These findings indicate that the selected sub-communities are heavily utilized for the dissemination of alternative news.
We also study the fraction of alternative news URLs with respect to overall news URLs (Figure~\ref{aggregate_month_counts-ratio}), highlighting that mainstream news URLs are overall more ``popular'' than the alternative news URLs.
Note that the Twitter spike in Figure~\ref{aggregate_month_counts-ratio} appears to be an artifact of a failure in our collection infrastructure.

\begin{figure}[t]
\centering
\subfigure[]{\includegraphics[width=0.37\textwidth]{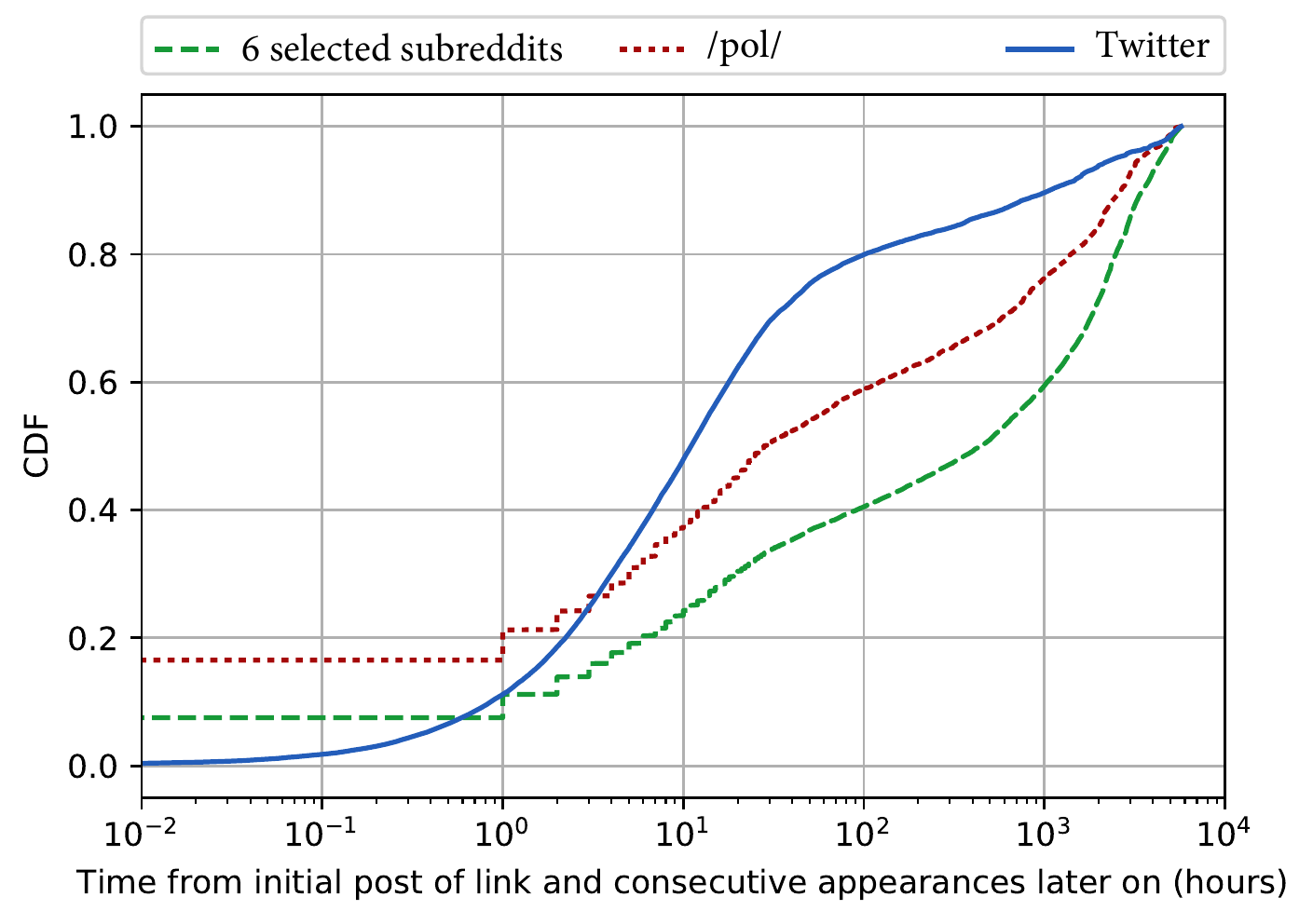}}\\[-1ex]
\subfigure[]{\includegraphics[width=0.37\textwidth]{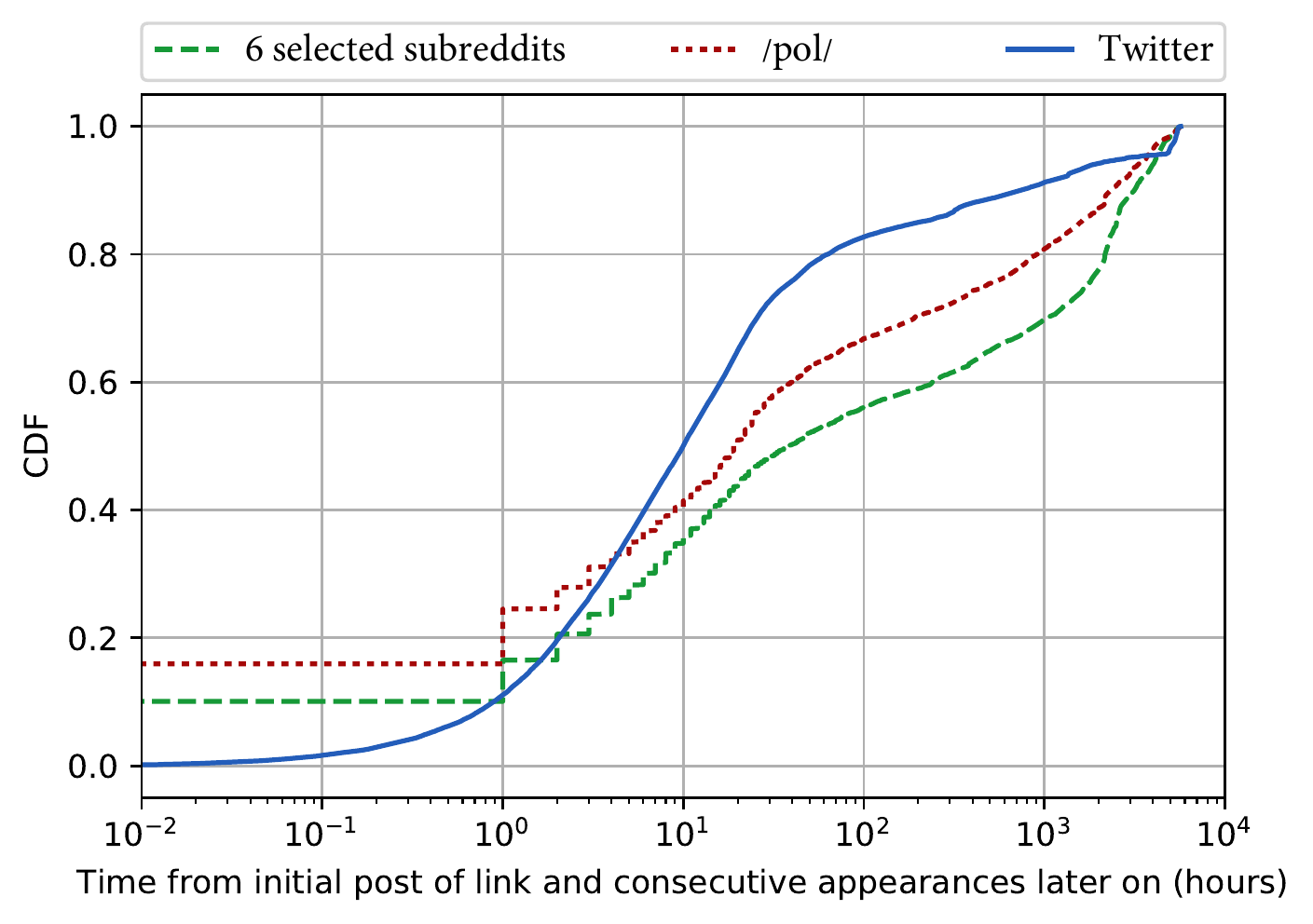}}
\caption{CDF of time difference (in hours) between the first occurrence of a URL and its next occurrences on each platform for (a) alternative and (b) mainstream news.}
\label{timeline_for_each_platform}
\end{figure}

As some users repost the same URL many times within the same platform,
we next study such reposting behavior and extract insights while comparing platforms.
In Figure~\ref{timeline_for_each_platform}, we plot the CDF of the time difference between the first occurrence of a URL and its next occurrences on the same platform.
Both alternative and mainstream news URLs are recycled over time within the platform (even after several months), but Twitter exhibits a smaller lag between the first occurrence and later ones compared to the other two platforms.
In all three platforms, there is an inflection point at the 24h period, which probably signifies the day-to-day behavior of news propagation within a platform, and this is true for both alternative and mainstream news.
Finally, mainstream news seem to propagate faster in these platforms than alternative news, especially on the six subreddits; for Twitter and \dspol the difference is not evident.
\begin{figure*}[t]
\subfigure[]{\includegraphics[width=0.37\textwidth]{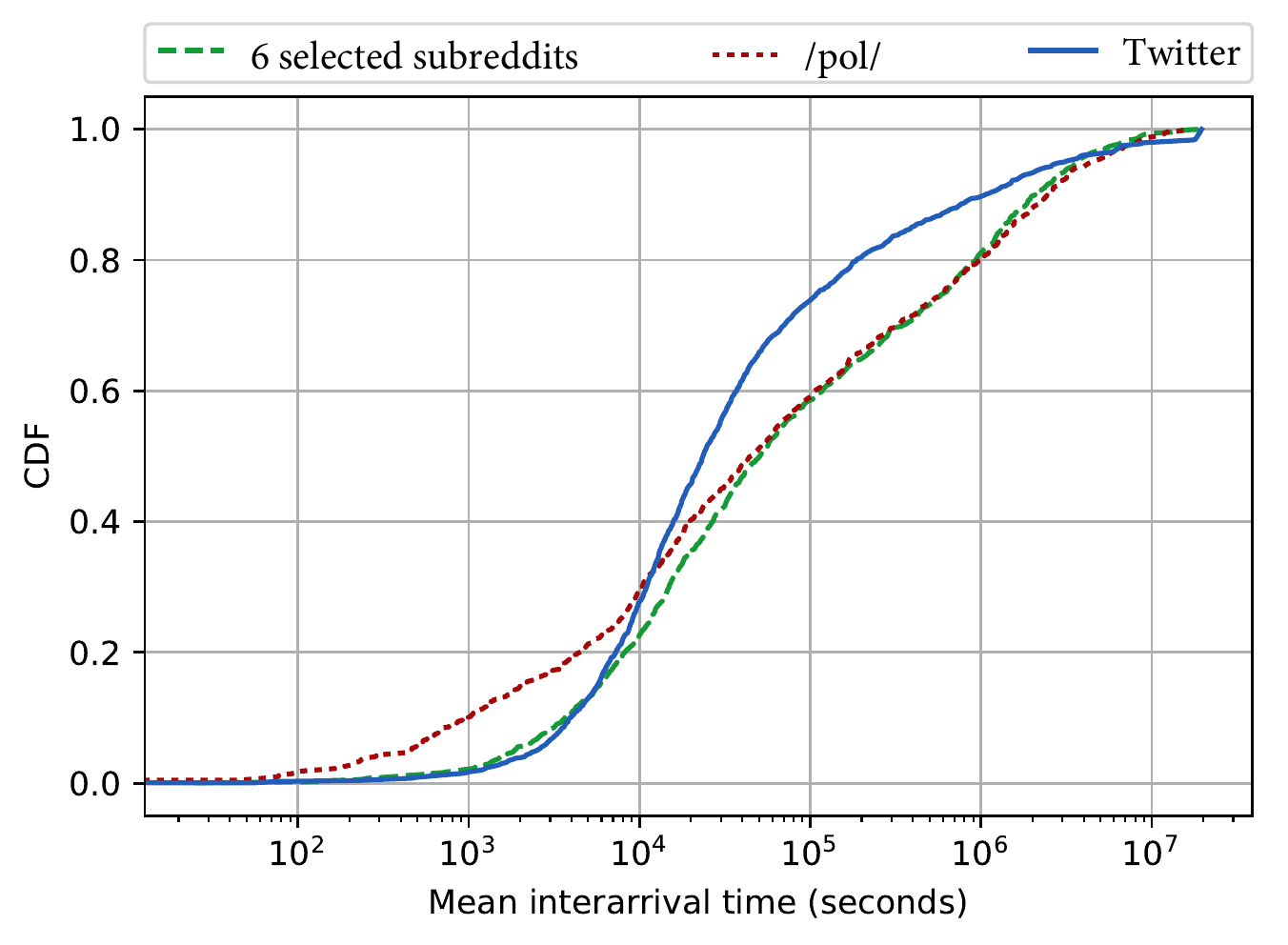}}
\hspace{0.2cm}
\subfigure[]{\includegraphics[width=0.37\textwidth]{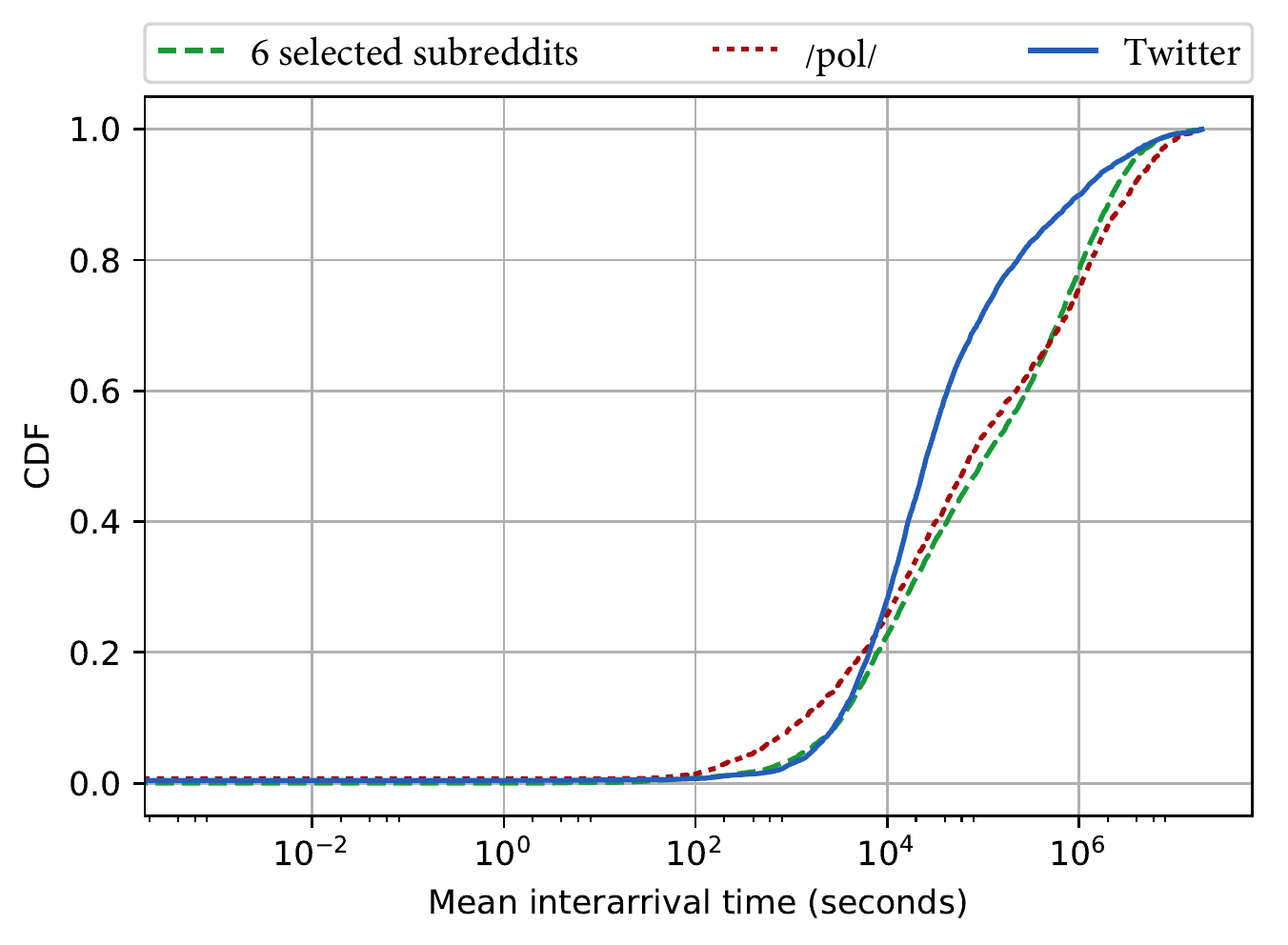}}
\subfigure[]{\includegraphics[width=0.37\textwidth]{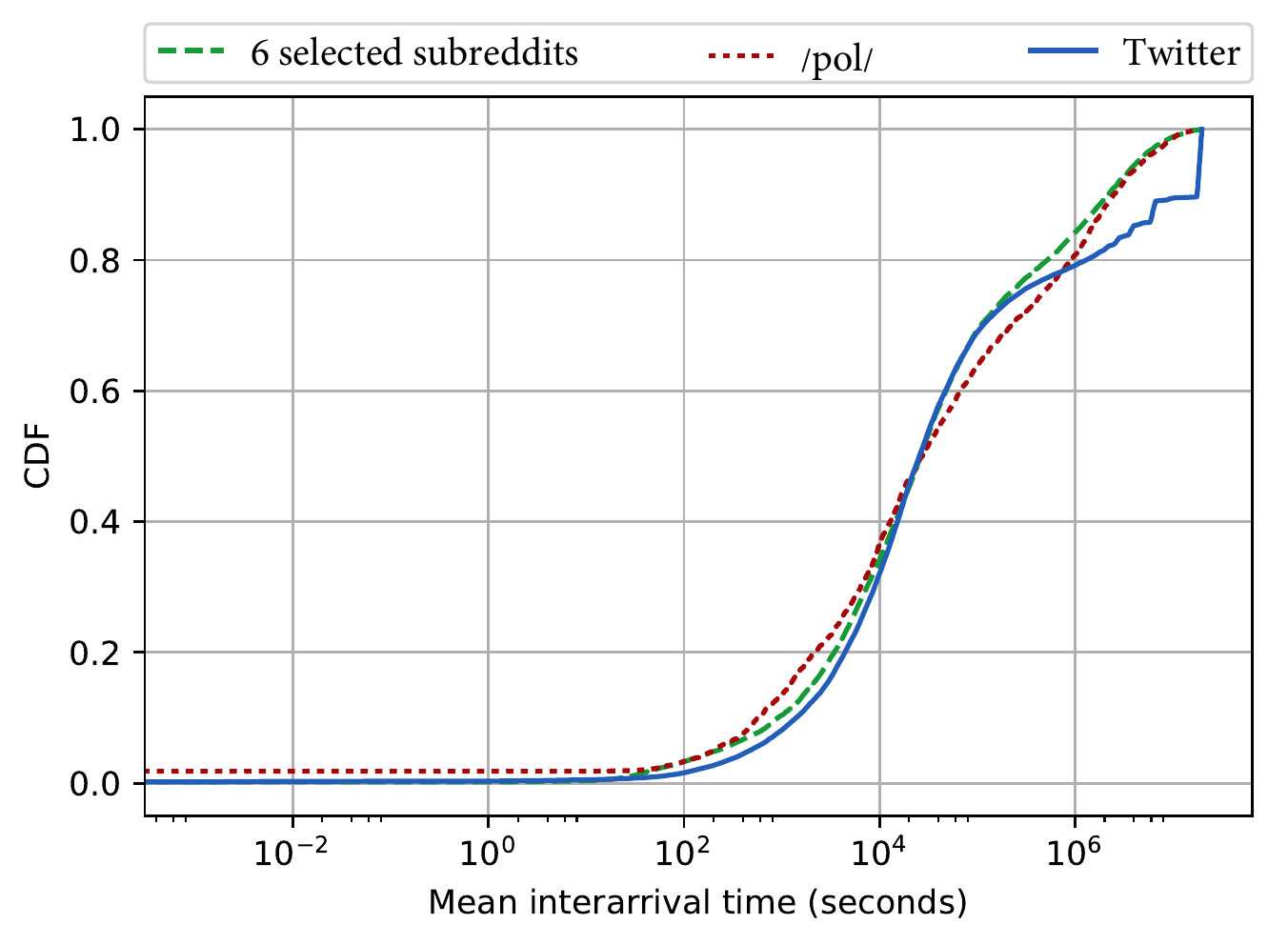}}
\hspace{0.2cm}
\subfigure[]{\includegraphics[width=0.37\textwidth]{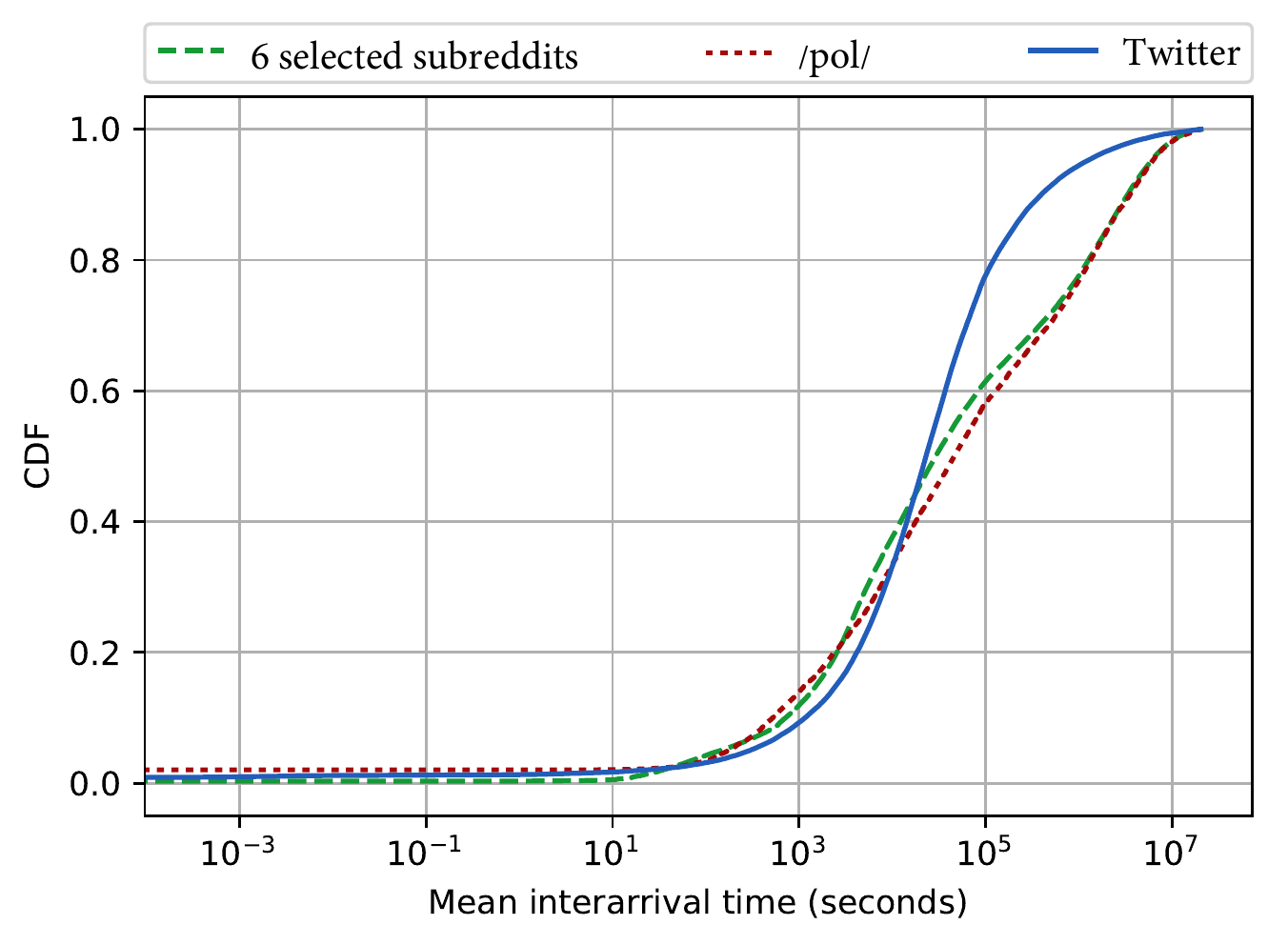}}
\caption{CDF for mean inter-arrival time for the URLs that occur more than once for (a) common alternative news URLs; (b) common mainstream news URLs; (c) all alternative news URLs, and (d) all mainstream news URLs.}
\label{cdf_interarrival_time}
\end{figure*}

We also study the inter-arrival time of reposted URLs.
Figure~\ref{cdf_interarrival_time} shows the CDF of the mean inter-arrival time of URLs that appear more than one time in each platform.
Each platform exhibits unique behavior, confirmed by a two sample Kolmogorov-Smirnov test showing significant differences between the distributions ($p < 0.01$ for each pairwise comparison).
However, \dspol and the six subreddits exhibit similar time-related sharing behavior for both mainstream and alternative news URLs, and Twitter has smaller mean inter-arrival time overall.
Interestingly, the six subreddits appear to have a duality in reposting behavior: for URLs with small inter-arrival time, it follows the faster pace of Twitter, whereas, for URLs with longer inter-arrival times, it follows \dspol.

\begin{table}[t]
\centering
\small
\begin{tabular}{@{}llrr@{}}
\hline
\textbf{Comparison} & \textbf{Type of  News} & \textbf{\#URLs where} & \textbf{\#URLs where} \\
{\bf (1st vs 2nd)} & & {\bf 1st is faster} & {\bf 2nd is faster}\\
 \hline
Reddit vs Twitter   & Mainstream         & 18,762                                                                                 & 11,416                                                                                 \\
                    & Alternative        & 5,232                                                                                  & 4,301                                                                                  \\ \hline
\dspol vs Twitter    & Mainstream         & 2,938                                                                                  & 4,700                                                                                  \\
                    & Alternative        & 778                                                                                    & 2,099                                                                                  \\ \hline
\dspol vs Reddit     & Mainstream         & 5,382                                                                                  & 14,662                                                                                 \\
                    & Alternative        & 1,455                                                                                  & 3,695                                                                                   \\ \hline
\end{tabular}
\caption{Statistics of URLs for the comparisons of time difference between platforms. Reddit refers to the six selected subreddits.}
\label{tbl:time_difference_statistics}
\end{table}

\subsection{Cross Platform Analysis}

\begin{figure*}[t!]
\subfigure[]{\includegraphics[width=0.337\textwidth]{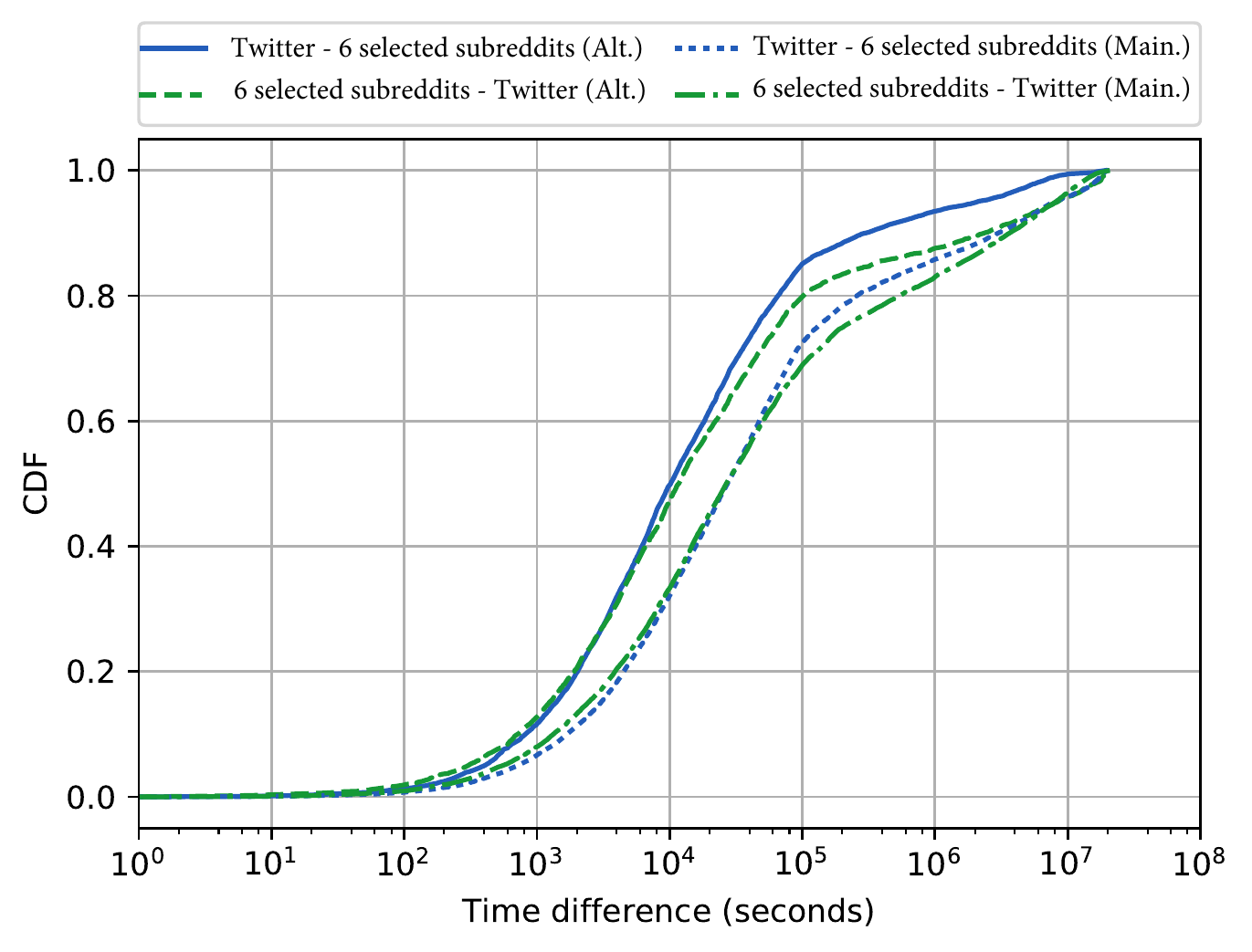}\label{cdf_time_difference-rt}}
\hspace{-0.3cm}
\subfigure[]{\includegraphics[width=0.337\textwidth]{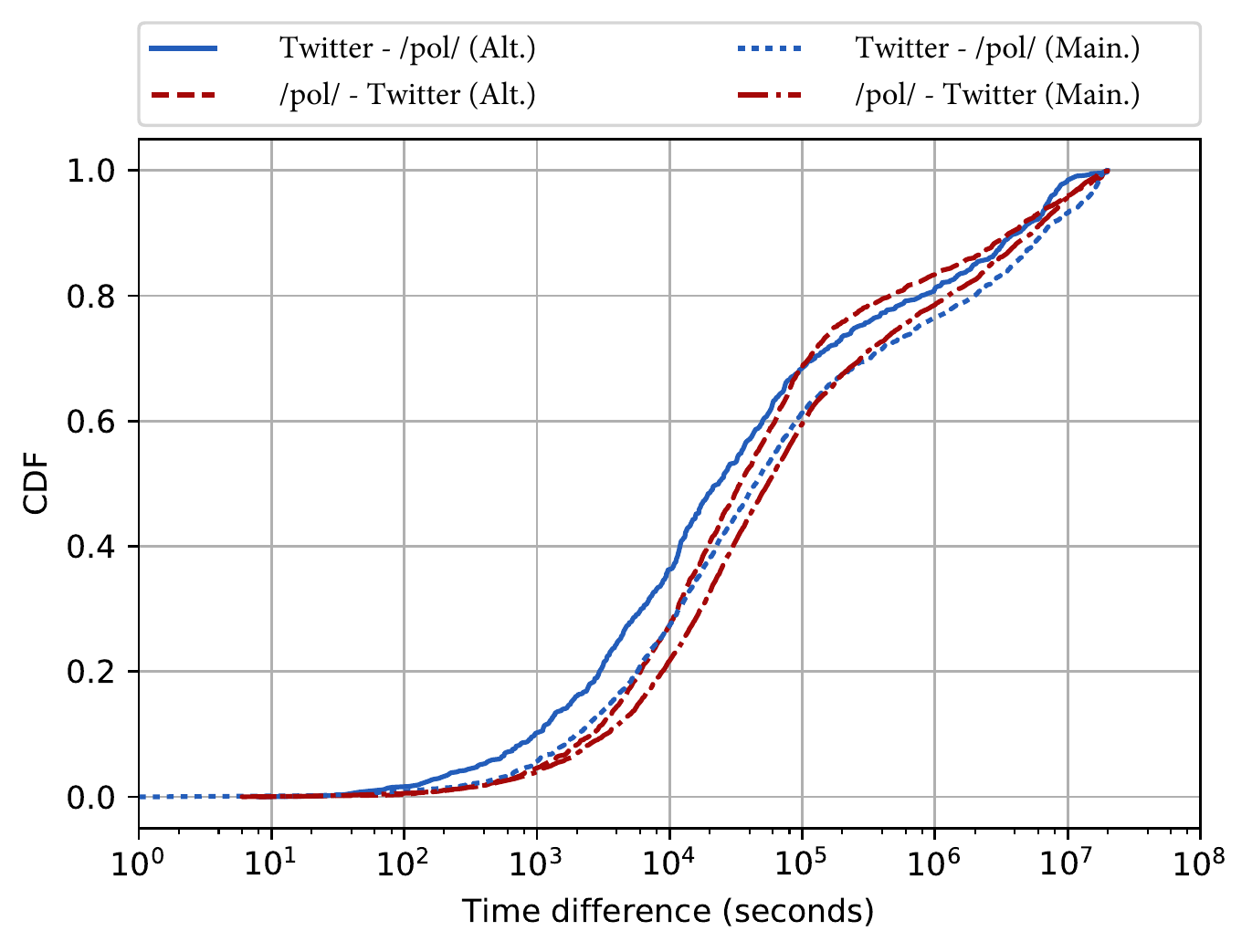}\label{cdf_time_difference-ct}}
\hspace{-0.3cm}
\subfigure[]{\includegraphics[width=0.337\textwidth]{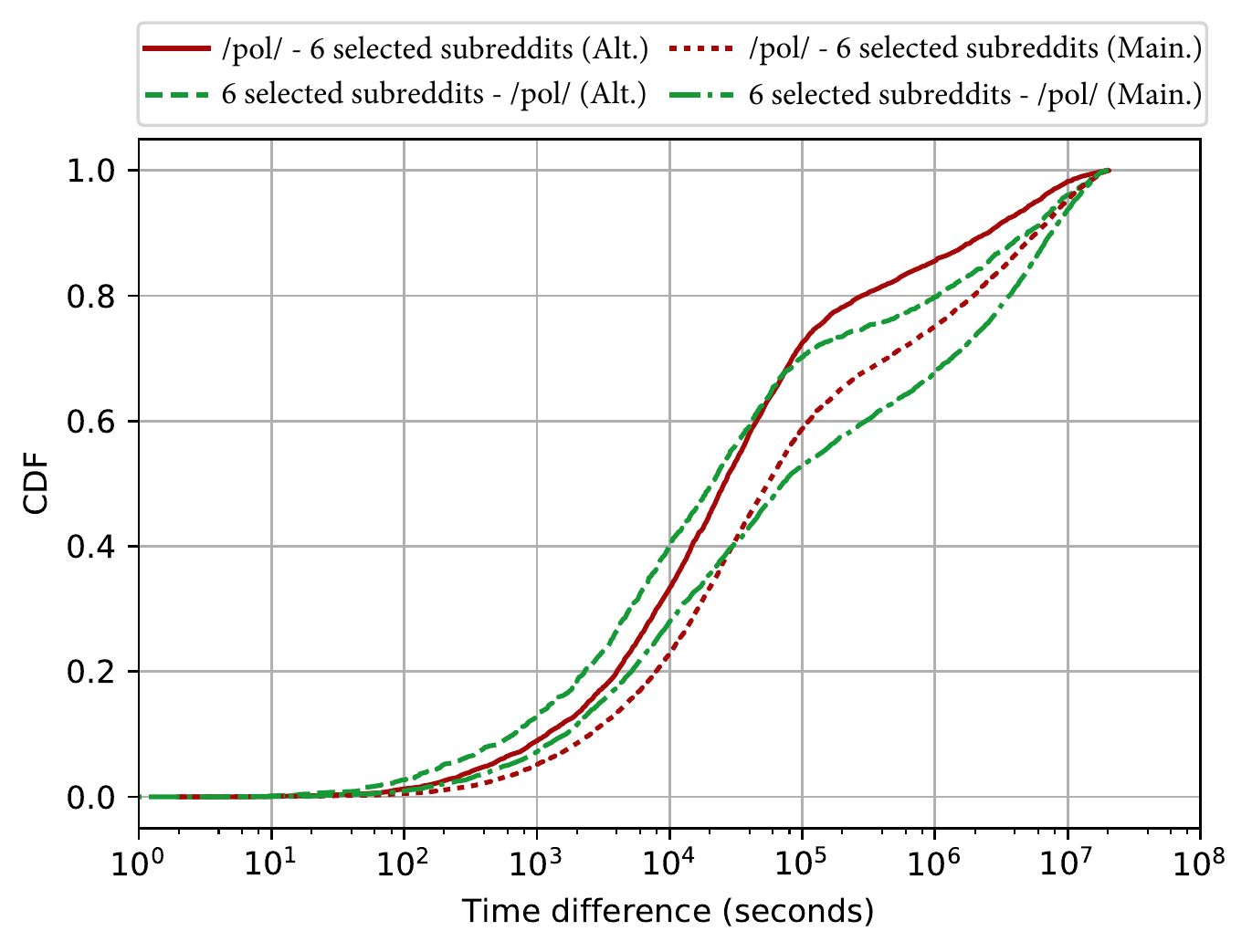}\label{cdf_time_difference-cr}}
\caption{CDF of the difference between the first occurrence of a URL between (a) six selected subreddits and Twitter, (b) /pol/ and Twitter, and (c) \dspol and six selected subreddits.}
\label{cdf_time_difference}
\end{figure*}

We now look at URLs that appear on more than one platform and study the time at which they are shared.
Figure~\ref{cdf_time_difference} plots the CDF of the time difference (in seconds) between the first occurrence of a
URL on pairs of platforms, while Table~\ref{tbl:time_difference_statistics} reports the numbers of URLs involved in each comparison.
We make the following observations: first, when comparing pairs of distributions for a given category of URLs, they are
statistically different (a two sample Kolmogorov-Smirnov test rejects the null hypothesis with $p<10^{-4}$).
Second, alternative news appear on multiple platforms faster than mainstream news.
This is consistent regardless of the pair of platforms we consider, and the sequence of appearances
(i.e., first in platform A and then B, vs. first in B and then in A).
Third, we notice the presence of a ``turning point'' with respect to the delay between URL appearance on
each platform, which seems to be consistent across all pairs of platforms and types of news, and matches the 24h period observed earlier.
Finally, there is a cross point when comparing URLs first posted on platform A and then on B, and URLs which were posted first in B and
then on A (i.e., when the lines for the same type of URLs cross).
Such a point represents which portion of URLs appear faster in one platform than the other.
For the Twitter-six selected subreddits comparison, alternative (mainstream) news appear faster on Twitter than
the six subreddits $80\%$ of the time ($50\%$), with these URLs exhibiting slower propagation, since the turning point is at $\sim$1 hour (5 hours).
Similarly, for the Twitter-\dspol comparison, alternative (mainstream) news appear faster on Twitter than \dspol $70\%$ ($5\%$) of the time,
with the turning point at 1 day (2 days).
Finally, for the six selected subreddits-\dspol comparison, alternative (mainstream) news appear faster on the six subreddits than \dspol for $65\%$ ($40\%$) of the time,
with the turning point around 18 hours (12 hours).

Next, given the set of unique URLs across all platforms and the time they appear for the first time, we analyze their appearance in one,
two, or three platforms, and the order in which this happens.
For each URL, we find the first occurrence on each platform and build corresponding ``sequences,'' e.g., if a URL first appears on the six  subreddits (Reddit) and subsequently on \dspol (4chan), the sequence is Reddit$\rightarrow$ 4chan (R$\rightarrow$4).
Table~\ref{links_sequences_all} reports the distribution of the sequences of appearances considering only the first hop, i.e.,
up to the first two platforms in the sequence.
The majority of URLs only appear on one platform: 82\% of alternative URLs and 89\% of mainstream news URLs.
Also, both alternative and mainstream news URLs tend to appear on the six subreddits first and later appear on either Twitter or \dspol, %
and
on Twitter before \dspol.

\begin{table}[t]
\centering
\small
\begin{tabular}{@{}lrrrr@{}}
\hline
{\bf Sequence} & \multicolumn{2}{r}{\bf Alternative (\%)} & \multicolumn{2}{r}{\bf Mainstream (\%)} \\ \hline
4 only       & 3,236 & (4.4\%)     & 18,654 & (3.7\%)   \\
4$\rightarrow$R      & 1,118 & (1.5\%)      & 4,606 & (0.9\%)    \\
4$\rightarrow$T      & 315 & (0.5\%)      & 861 & (0.17\%)     \\ \hline
R only       & 24,292 & (33.3\%)    & 230,602 & (46.1\%)  \\
R$\rightarrow$4      & 2,181 & (3.0\%)     & 11,307 & (2.3\%)    \\
R$\rightarrow$T      & 4,769 & (6.5\%)     & 16,685 & (3.35\%)   \\ \hline
T only       & 32,443 & (44.5\%)    & 204,836 & (41\%)   \\
T$\rightarrow$4      & 585 & (0.8\%)      & 1,345 & (0.26\%)     \\
T$\rightarrow$R      & 3,964 & (5.5\%)     & 10,640 & (2.12\%)    \\ \hline
\end{tabular}
\caption{Distribution of URLs according to the sequence of first appearance within platforms for all URLs, considering only the first hop. ``4'' stands for \dspol (4chan), ``R'' for the six selected subreddits (Reddit), and ``T'' for Twitter.}
\label{links_sequences_all}
\end{table}

We also study the temporal dynamics of URLs that appear on all three platforms, with triplets of sequences.
Table~\ref{links_sequences_common} reports the distribution of these sequences.
The most common sequences are similar for both alternative and mainstream news URLs:
R$\rightarrow$T$\rightarrow$4, R$\rightarrow$4$\rightarrow$T, and T$\rightarrow$R$\rightarrow$4 are the top three sequences.
As already mentioned, the six selected subreddits ``outperform'' both other platforms in terms of the speed of sharing mainstream and alternative news URLs, as evidenced by the fact that it is at the head of the sequence for 51\% and 59\% of alternative and mainstream news URLs, respectively.

\begin{table}[t]
\centering
\small
\begin{tabular}{@{}lrrrr@{}}
\hline
{\bf Sequence} & \multicolumn{2}{r}{\bf Alternative (\%)} & \multicolumn{2}{r}{\bf Mainstream (\%)} \\ \hline
4$\rightarrow$R$\rightarrow$T    & \hspace*{0.2cm} 128 & (5.5\%)      & 552 & (8.9\%)     \\
4$\rightarrow$T$\rightarrow$R    & 145 & (6.2\%)      & 290 & (4.7\%)     \\
R$\rightarrow$4$\rightarrow$T    & 335 & (14.4\%)      & 1,525 & (24.5\%)    \\
R$\rightarrow$T$\rightarrow$4    & 841 & (36.3\%)      & 2,189 & (35.3\%)    \\
T$\rightarrow$4$\rightarrow$R    & 192 & (8.2\%)      & 486 & (7.8\%)     \\
T$\rightarrow$R$\rightarrow$4    & 673 & (29\%)       & 1,166 & (18.8\%)    \\ \hline
\end{tabular}
\caption{Distribution of URLs according to the sequence of first appearance within a platform for URLs common
to all platforms. ``4'' stands for \dspol (4chan), ``R'' for the six selected subreddits (Reddit), and ``T'' for Twitter.}
\label{links_sequences_common}
\end{table}

\begin{figure*}[t!]
\subfigure[]{\includegraphics[width=0.435\textwidth]{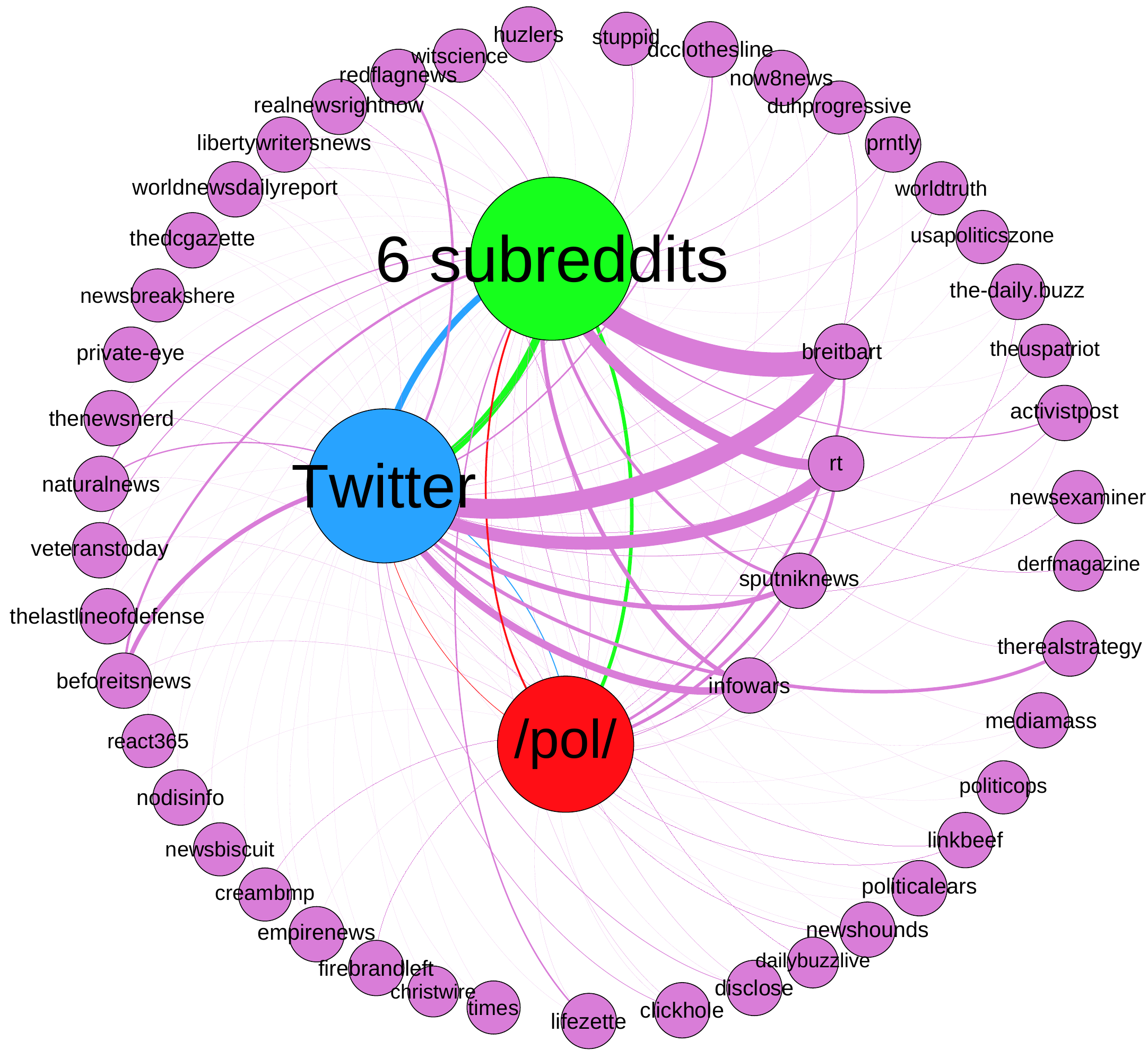}\label{timeline_all_fake-a}}
\hspace{0.2cm}
\subfigure[]{\includegraphics[width=0.435\textwidth]{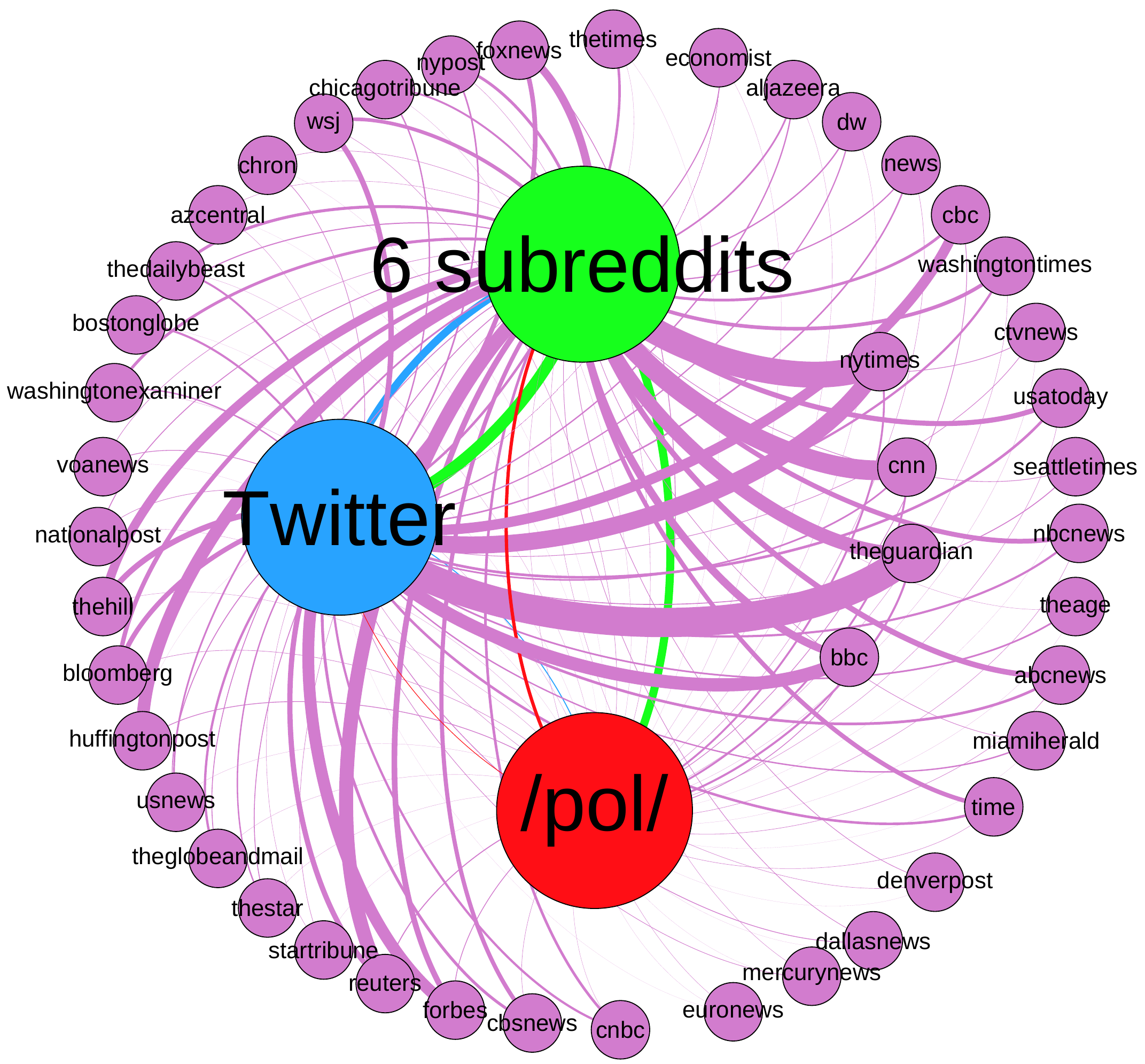}\label{timeline_all_fake-b}}
\caption{Graph representation of news ecosystem (a) alternative news domains and (b) mainstream news domains. Edges are colored the same as their source node.}
\label{timeline_all_fake}
\end{figure*}

Finally, we analyze the source of the URLs for each of the three platforms, as follows.
We create two directed graphs, one for each type of news, $\boldsymbol{G}=(\boldsymbol{V},\boldsymbol{E})$, where
$\boldsymbol{V}$ represents alternative or mainstream domains, as well as the three platforms, and $\boldsymbol{E}$ the set
of sequences that consider only the first-hop of the platforms.
For example, if a \url{breitbart.com} URL appears first on Twitter and later on the six selected subreddits, we add an edge
from \url{breitbart.com} to Twitter, and from Twitter to the six selected subreddits.
We also add weights on these edges based on the number of such unique URLs.
By examining the paths %
we can discern which domains' URLs tend to appear first on each of the platforms.
Figure~\ref{timeline_all_fake} shows the graphs built for alternative and mainstream domains.
Comparing the thickness of the outgoing edges, one can see that \url{breitbart.com} URLs appear first in
the six selected subreddits more often than on Twitter and more frequently than on \dspol.
However, for other popular alternative domains, such as \url{infowars.com}, \url{rt.com}, and \url{sputniknews.com}, URLs
appear first on Twitter more often than the six selected subreddits and \dspol. Also, \dspol is rarely the platform where a URL first shows up.
For the mainstream news domains, we note that URLs from \url{nytimes.com} and \url{cnn.com} tend to appear first more often on the selected subreddits
than Twitter and \dspol,
however, URLs from other domains like \url{bbc.com} and \url{theguardian.com} tend to appear first more often on Twitter than the selected subreddits.
Similar to the alternative domains graph, there is no domain where \dspol dominates in terms of first URL appearance.
\subsection{Take-Aways} 
In summary, our temporal dynamics analysis shows that:
1) Specific sub-communities are heavily utilized for the dissemination of alternative news (Figure~\ref{aggregate_month_counts});
2) Twitter URL posting behavior is more aggressive for both alternative and mainstream news when compared to the other two platforms (Figure~\ref{timeline_for_each_platform} and ~\ref{cdf_interarrival_time});
3) Both alternative and mainstream URLs tend to appear on the six selected subreddits before appearing on Twitter and/or \dspol (Table~\ref{tbl:time_difference_statistics} and Figure~\ref{cdf_time_difference}).
This may be due the nature of the Reddit platform, where URL posting is the prevalent way to disseminate information;
4) URLs from specific domains tend to appear first on a particular platform.
For example, for alternative news breitbart.com, URLs appear first on the six selected subreddits, while on Twitter the same applies for infowars.com, rt.com, and sputniknews.com.
The \dspol does not dominate for any of the domains (Figure~\ref{timeline_all_fake}).

Note that our analysis does not factor out %
the gaps in our Twitter dataset.
Consequently, some of the results may be affected by the fact that some URLs from Twitter might be missing.
To address this issue, in Section~\ref{sec:influence_measurement}, we omit a subset of the URLs impacted by the gaps in our datasets.
Furthermore, the use of a rigorous statistical model (Hawkes processes) helps to mitigate this potential issue. %

\section{Influence Estimation}
\label{sec:influence_measurement}

Thus far, our measurements have shown relative differences in how news media is shared on Reddit, Twitter, and 4chan.
In this section, we provide meaningful evidence of how the individual platforms influence the media shared on other platforms.
We do so by using a mathematical technique known as Hawkes processes.
These statistical models can be used for modeling the dissemination of information in Web communities~\cite{farajtabar2017fake} as well as measuring social influence~\cite{guo2015bayesian}.

\begin{figure}[t]
\includegraphics[width=\columnwidth]{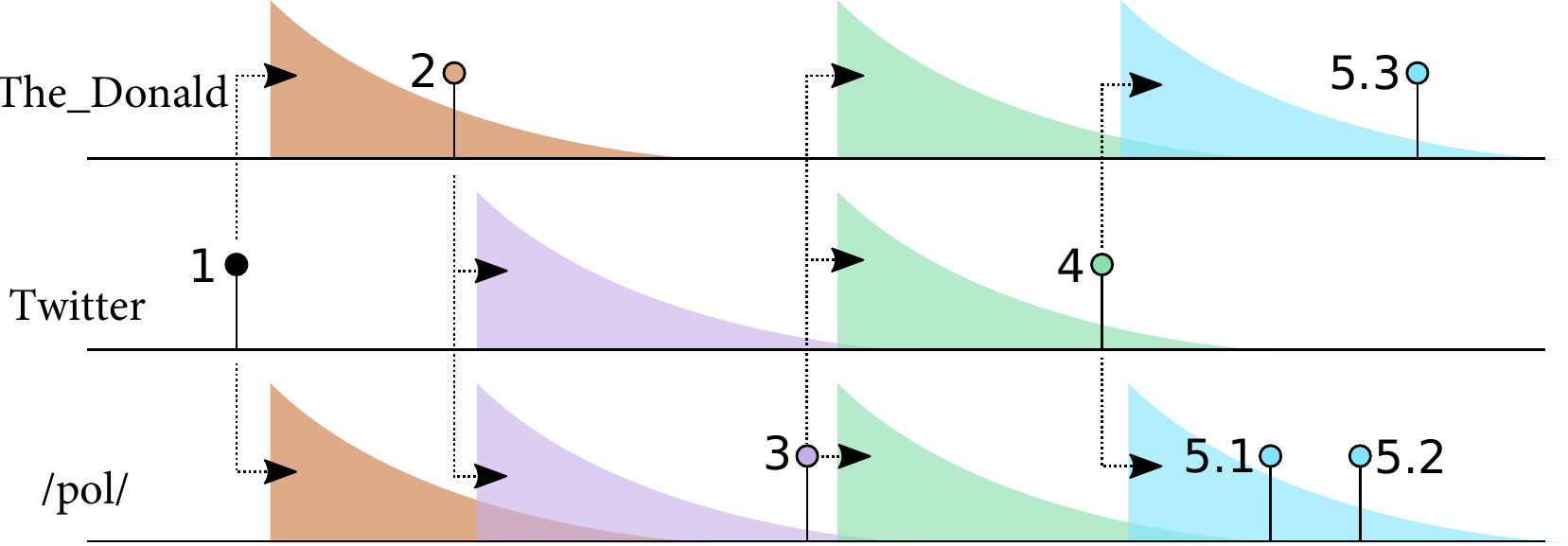}
\caption{A depiction of a Hawkes model showing how the interaction between events on 3 processes might look like.
Note: this example is inspired by~\cite{linderman2014}.}
\label{fig:hawkes}
\end{figure}

\subsection{Hawkes Processes}

Note that the three platforms we measure do not obviously exist in a vacuum, rather, they exist within the greater ecosystem of the Web.
Imagine, however, that each of the platforms were entirely self-contained, with a completely disjoint set of users.
In such a scenario, there would be a natural rate at which URLs will be posted, and it would be possible to model
this using standard Poisson processes.
However, our platforms are clearly \emph{not} independent.
While they do exhibit their own {\it background} URL posting rates and internal influence, they are also affected by each other, as well as by the greater Web.

A Hawkes model consists of a number, $K$, of point processes, each with a ``background rate'' of events $\lambda_{0,k}$.
An event on one process can cause an impulse response on other processes, which increases the probability of an event occurring above the processes' background rates.
Figure~\ref{fig:hawkes} depicts an example of what a sequence of events on a Hawkes model with three processes might look like, using The\_Donald, Twitter, and \dspol communities for representative purposes.

First, an event $1$ (a URL being posted) occurs on Twitter; this is caused by the background rate of the process, meaning that the URL was posted not because it was seen on any of the platforms in the model, but because it was seen elsewhere (including a user finding it organically).
This initial event causes an impulse response on the rates of the other processes, The\_Donald and \dspol, meaning that the URL is more likely to be posted on those platforms after having been seen on Twitter.
Eventually, this causes another event on The\_Donald ($2$), which in turn causes an event on \dspol.
A process can cause an additional impulse response to itself, as seen with event 3, and multiple events can be caused
in response to a single event, as seen with event 4 causing events 5.1, 5.2, and 5.3.
Naturally, the data we collect does not explicitly state which events are caused by other events, or which are caused by the background rate.

For our purposes, fitting a Hawkes model to a series of events on the different platforms gives us values for the background rates for each platform along with the probability of an event on one platform causing events on other platforms.
We emphasize that the background rates of the Hawkes processes allows us to also account for the probability of an event caused by external sources of information. 
For example, it includes the probabilities of events caused by Facebook as well as other platforms.
Thus, while we are only able to model the specific influences of the three platforms we study, the resulting probabilities are affirmatively attributable to each of them; the influence of the greater Web is captured by the background rates.

For a discrete-time Hawkes model, time is divided into a series of bins of duration $ \Delta t$, and events occurring
within the same time bin do not interact with each other.
The rate of each $k$-th process,
$\lambda_{t,k}$ is given by:
\[
  \lambda_{t,k} = \lambda_{0,k} + \sum_{k'=1}^K \sum_{t'=1}^{t-1} s_{t',k'} \cdot h_{k' \to k}[t - t']
\]
where $ s \in \mathbb{N}^{T \times K}$ is the matrix of event counts (how many events occur for process $k$ at time $t$) and $h_{k' \to k}[ t-t']$ is an impulse response function that describes the amplitude of influence that events on process $k'$ have on the rate of process $k$.

Following~\cite{linderman2014}, the impulse response function $h_{k \to k'}[t-t']$ can be decomposed into a scalar weight $W_{k \to k'}$ and a probability mass function $G_{k \to k'}[d]$.
The weight specifies the strength of the interaction from process $k$ to process $k'$ and the probability mass function specifies how the interaction changes over time:
\[
  h_{k \to k'}[d] = W_{k \to k'} G_{k \to k'}[d]
\]
The weight value $W_{k \to k'}$ can be interpreted as the expected number of child events that will be caused on process $k'$ after an event on process $k$.
The probability mass function $G_{k \to k'}$ specifies the probability that a child event will occur at each specific time lag $d \Delta t $, up to a maximum lag $\Delta t_{max}$.
This interpretation of $W_{k \to k'}$ is useful because it allows us to compare how much influence platforms have on each other.
For instance, we can examine whether a URL posted on Twitter or on Reddit is more likely to cause the same URL to be posted on 4chan, or if there is a difference in influence from one platform to another between URLs for mainstream and alternative news sites.

\subsection{Methodology}
We now provide more details about our experiments, once again, considering 4chan (\dspol), Twitter, and the six subreddits.
We study Hawkes processes at the subreddit granularity to get a better understanding of the various platforms and particular subreddits.

We aim to examine how these platforms and subreddits influence each other, so we model the arrival of URLs, in posts or tweets, with a Hawkes model with $K=8$ point processes---one for Twitter, one for \dspol, and one for each of the subreddits.
The model is fully connected, i.e., it is possible for each process to influence all the others, as well as itself, which describes behavior where participants on a platform see a URL and re-post it on the same platform.
For example, with Twitter, this value ($W_{\mathrm{Twitter} \to \mathrm{Twitter}}$) would likely be quite high, given that tweets are commonly re-tweeted a number of times: the initial tweet containing a URL is likely to cause a number of re-tweets, also containing the URL, on the same platform.

We select URLs that have at least one event in Twitter, \dspol, and at least one of the subreddits, and we model each URL individually.
The missing Twitter data (due to our dataset gaps; see Section~\ref{sec:data_collection}) affects 3,177 (37\%) of the URLs.
One way to mitigate the impact of this missing data is to remove events for which it has a larger impact.
E.g., if an event spans 100 days, the missing Twitter data has less of an effect than if the event only spanned two days.
Thus, we examine URLs from  other platforms that overlap with any of the missing days and remove the 10\% of URLs (895) with the shortest total duration from the first event recorded until the last event recorded.
This results in the missing data making up a smaller portion of the overall duration of the events.

\begin{table*}[t]
\centering

\small
  \begin{tabular}{llrrrrrrrr}
  \hline
      &        &     {\bf The\_Donald} &  {\bf worldnews} &   {\bf politics} &  {\bf news} &   {\bf conspiracy} &   {\bf AskReddit} &  {\bf \dspol} &  {\bf Twitter} \\
  \hline
URLs  & Mainstream &  3,097 &   2,523 &  3,578 &  2,584 &   907 &  841 &   5,589 &     5,589 \\
      & Alternative &  2,008 &    252 &   813 &   362 &   321 &  100 &   2,136 &     2,136 \\
      &   Total &  5,105 &   2,775 &  4,391 &  2,946 &  1,228 &  941 &   7,725 &     7,725 \\

  \hline

Events & Mainstream &  12,312 &   7,517 &  26,160 &  5,794 &  1,995 &  2,302 &  19,746 &    36,250 \\
       & Alternative &   7,797 &    458 &   2,484 &   586 &   497 &   176 &   7,322 &    23,172 \\
       &Total &  20,109 &   7,975 &  28,644 &  6,380 &  2,492 &  2,478 &  27,068 &    59,422 \\

\hline
Mean $\lambda_0$  & Mainstream &  0.001502 &  0.001382 &  0.001265 &  0.001392 &  0.000501 &  0.000107 &  0.001564 &  0.002330 \\
                  & Alternative &  0.001627 &  0.000619 &  0.000696 &  0.000553 &  0.000423 &  0.000034 &  0.001525 &  0.002803 \\
\hline
\end{tabular}
\caption{Total URLs with at least one event in Twitter, \dspol, and at least one of the subreddits; total events for mainstream and alternative URLs, and the mean background rate ($\lambda_0$) for each platform/subreddit. Note that the numbers reported on this table are lower than the ones reported by Tables~\ref{tbl:datasets} and ~\ref{table-tweets-stats} due to the fact that we only select URLs that appear in multiple sub-communities within the three platforms and we drop a number of URLs that
overlap the time period where data is missing.}
\label{tbl:hawkes}
\end{table*}

The number of remaining URLs and events included for each platform are shown in Table~\ref{tbl:hawkes}.
 For each URL, we create a matrix $s \in \mathbb{N}^{T \times 8}$ containing the number of events (URL posts) per minute for each of the  platforms/subreddits.
Here, $T$ is the number of minutes from the first recorded post of the URL on any platform, to the last recorded post of a URL on any platform (\textbf{NB:} this value can be different for each URL).
We select $ \Delta t = 1~\mathrm{minute}$ as a reasonable compromise between accuracy and computational cost.
Using this bin size, 92\% of events are in a bin by themselves, and another 5.4\% share a bin, but only with other events from the same platform or subreddit, meaning that timing interactions between the platforms are not lost.

Next, we fit a Hawkes model for each URL using the approach described in~\cite{linderman2014,lindermanArxiv}, which uses Gibbs sampling to infer the parameters of the model from the data, including the weights, background rates, and shape of the impulse response functions between the different processes.
By setting $\Delta t_{max} = 60 \cdot 12 = 720$ minutes, we say that a given event can cause other events within a 12-hour time window.
Experiments with other values (6, 12, 24, and 48 hours) gave similar results.
After fitting the models, we have the values for the $W$ matrix -- i.e., the weights of the interactions between events on different processes for each URL.
These weights can then be interpreted as the expected number of events.
For example, $W_{Twitter \to /pol/} = 0.1$ would mean that an event on Twitter will cause $n$ events on \dspol, where $n$ is drawn from a Poisson distribution with rate parameter $0.1$.
Finally, for each URL, we also get the $\lambda_{0,k}$ values for each process, which are the background rates for event arrivals that are \emph{not} caused by other events in the system we model.
Again, these background rates capture events due to some \emph{other} platform, e.g., someone posting the URL after reading it on the original site or seeing the URL on another site not included in the model, like Facebook.

\subsection{Results}

\begin{figure}[t]
\hspace*{-0.1cm}\includegraphics[width=1.02\columnwidth]{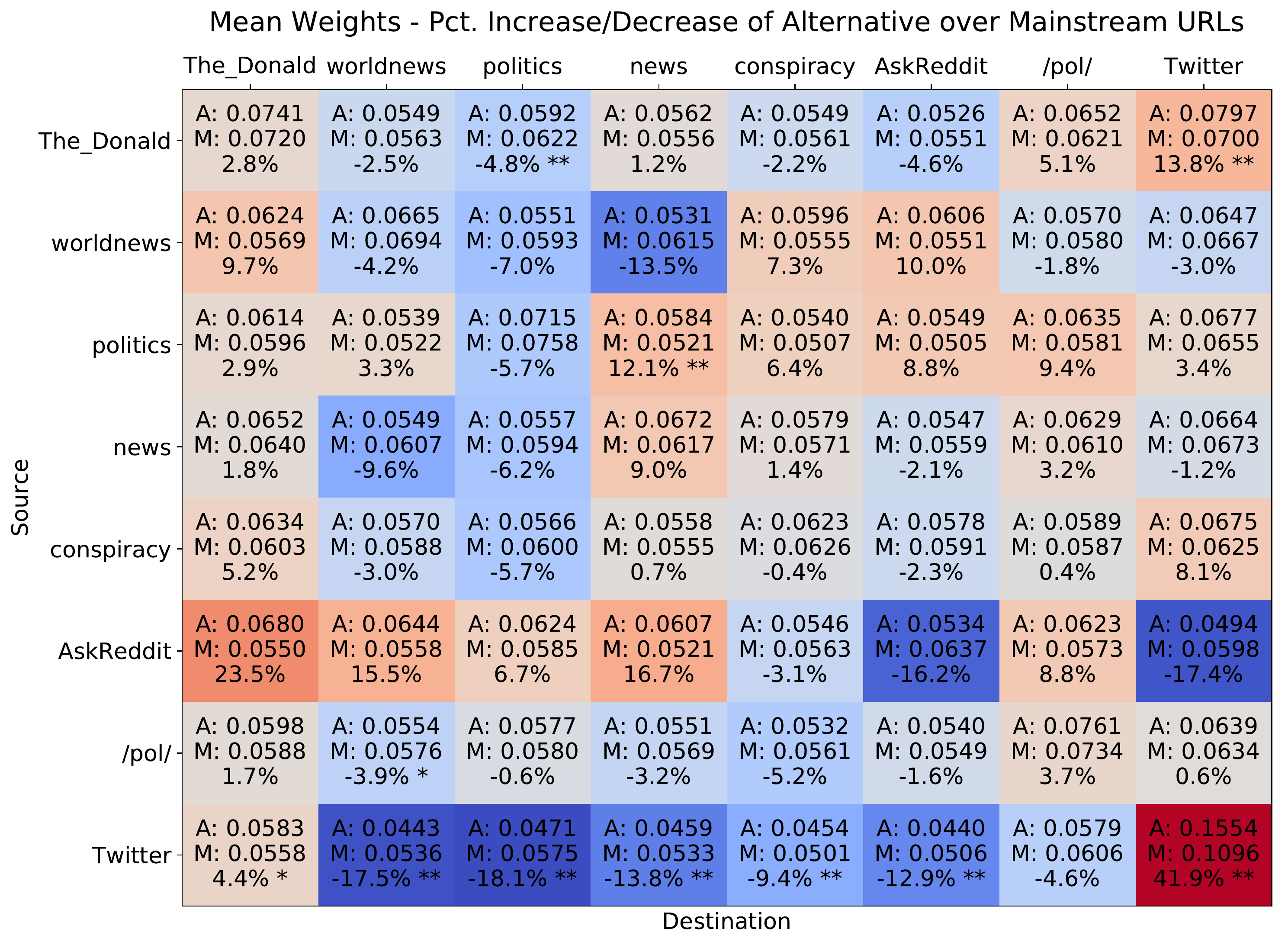}
\caption{Mean weights for alternative URLs (A), mainstream URLs (M), and the percent increase/decrease between mainstream and alternative (also indicated by the coloration). 
Stars indicate the level of statistical significance (p-value) between the weight distributions: no stars indicate no statistical significance, while * and ** indicate, resp., statistical significance with $p < 0.05$ and $p < 0.01$. 
}
\label{fig:mean-weights-difference}
\end{figure}

Looking at the number of URLs in Table~\ref{tbl:hawkes}, we note that there are substantially more events for mainstream than alternative news URLs.
However, for Twitter, \dspol, and The\_Donald, the ratios of events to URLs for alternative news URLs are similar to or greater than the ratios for mainstream ones.
These high ratios explain the high background rates (cf.~Table~\ref{tbl:hawkes}) for alternative news URLs for these platforms despite the lower number of events.

From the Hawkes models for each URL, we obtain the weight matrix $W$ which specifies the strength of the connections between the different platforms and subreddits.
The mean weight values over all URLs for alternative and mainstream news URLs, as well as the percentage difference between them are presented in Figure~\ref{fig:mean-weights-difference}.
First, we look at Twitter.
Background rates are high for both mainstream and alternative news URLs, which is not surprising given the large number of users on the platform.
The values for $W_{\mathrm{Twitter} \to \mathrm{Twitter}}$ are also substantially higher than all other weights: $0.1096$ for mainstream news URLs and $0.1554$ for alternative news URLs.
This reflects the ease and common practice of re-tweeting: a URL in a tweet is likely to generate other events as users re-tweet it.
There are different possible explanations for why the Twitter to Twitter rate for alternative news URLs is much greater than the rate for mainstream news URLs.
The first is bot activity---if automated Twitter bots are used to spread alternative news URLs, it could result in a much higher rate of tweeting and re-tweeting.
Another possible explanation is the behavior of users who read news stories from alternative sources; they might be more inclined to re-tweet the URL~\cite{gupta2013faking}.

Looking at the weights for Twitter to the other platforms, except The\_Donald, they are all greater for mainstream news URLs, meaning that the average tweet containing a mainstream URL is more likely to cause a subsequent post on the other platforms than the average tweet containing an alternative URL.
The next communities most likely to cause events on others are The\_Donald and \dspol.
It is worth noting that The\_Donald is the only platform/subreddit that has greater alternative URL weights for all of its inputs.
Assuming that the population of The\_Donald users that also read, say, worldnews is the same for both alternative and mainstream news URLs---which is reasonable---then the difference in weights implies that the users have a stronger preference for re-posting alternative news URLs back to The\_Donald than for mainstream news URLs.
The opposite can be seen for worldnews and politics, where most of the input weights are stronger for mainstream news.
However, despite the higher weights for alternative news URLs, The\_Donald is also, interestingly, influenced more strongly by mainstream news URLs than alternative news URLs on all platforms, with the exception of Twitter.  This is in part because of the greater number of mainstream URL events, but The\_Donald also has a higher background rate for alternative news URLs than mainstream news URLs, which implies that a lot of the alternative news URLs on the platform are coming from other sources.

\begin{figure}[t]
\hspace*{-0.1cm}\includegraphics[width=1.02\columnwidth]{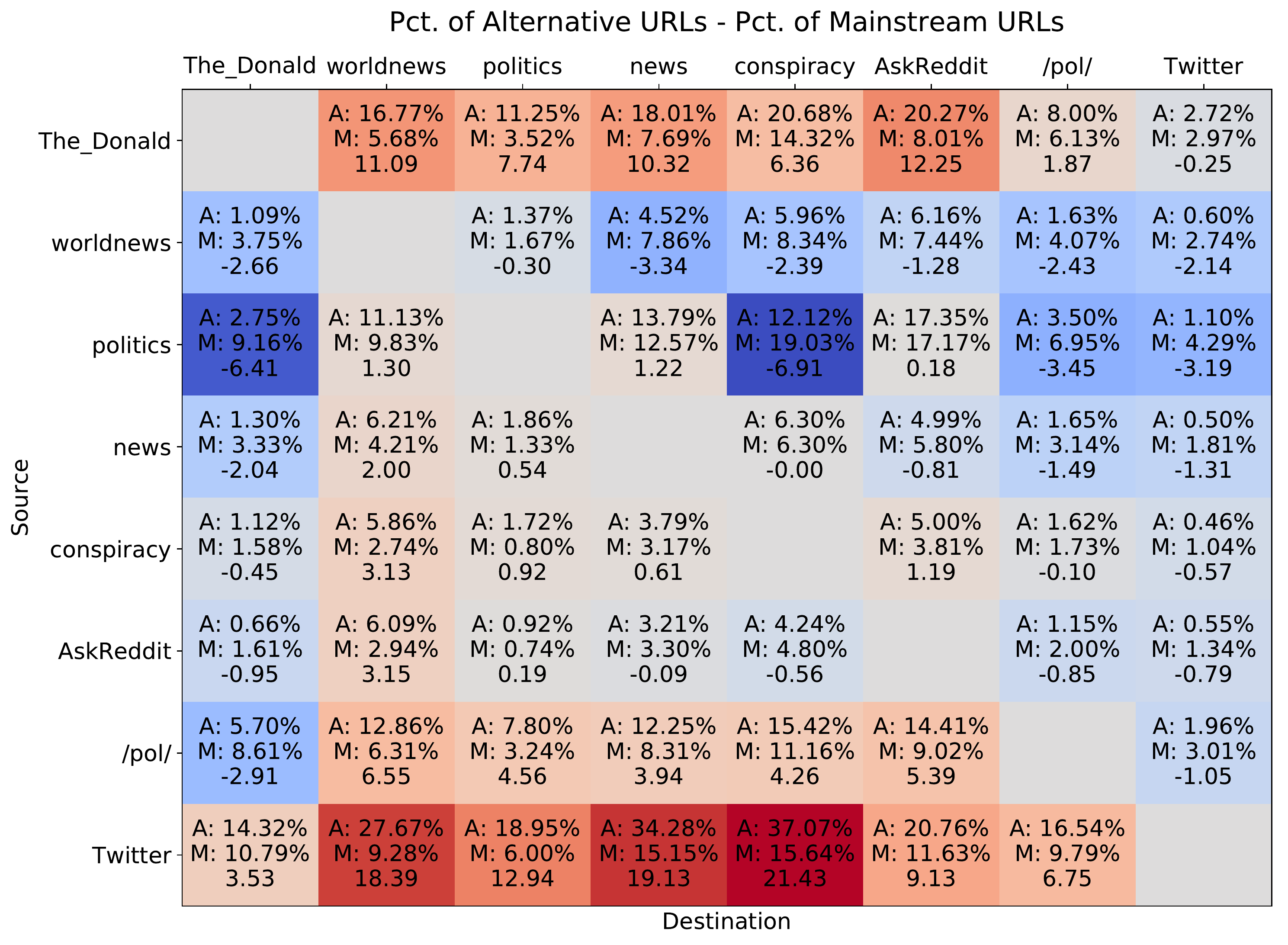}
\caption{Mean estimated percentage of alternative URL events caused by alternative news URL events (A), mean estimated percentage of mainstream news URL events caused by mainstream news URL events (M), and the difference between alternative and mainstream news (also indicated by the coloration).
}
\label{fig:percent-events-difference}
\vspace{-0.2cm}
\end{figure}

To assess the statistical significance of the results, we perform two sample Kolmogorov-Smirnov tests on the weight distributions of mainstream and alternative news URLs for each source-destination pair (depicted as stars in Figure~\ref{fig:mean-weights-difference}).
Rejecting the null hypothesis here implies a difference in the way mainstream and alternative news URLs propagate from the source to the destination---either mainstream news URLs tend to cause more events on other platforms than alternative news URLs, or the opposite.
Unsurprisingly, many of the source-destination pairs have no significant difference. 
However, in most cases where Twitter is the source community there \emph{is} a significant statistical difference with $p < 0.01$.
I.e., for some communities, Twitter is used not just to disseminate news, but to disseminate news from a specific \emph{type} of source.

Figure~\ref{fig:percent-events-difference} illustrates the estimated total impact of the different platforms on each other, for both mainstream and alternative news URLs.
The impact is estimated using the weight values that are shown in Figure~\ref{fig:mean-weights-difference}. Since the weight values can be interpreted as the expected number of \emph{additional} events caused as a consequence of an event, we can estimate the percentage of events on each platform that were caused by each of the other platforms by multiplying the weight by the actual number of events that occurred on the source platform and dividing by the number of events that occurred on the destination platform: %
\[
  \mathrm{Pct}_{A \to B} = \frac{ \sum_{u \in \mathrm{urls}} \left( W_{A \to B} \cdot \sum_{t = 1}^T s_{t,A} \right) }{  \sum_{u \in \mathrm{urls}} \sum_{t = 1}^T s_{t,B} } %
\]
The percentages for mainstream and alternative news URLs as well as the difference between them are presented in Figure~\ref{fig:percent-events-difference}.

Twitter contributes heavily to both types of events on the other platforms---and is in fact the most influential single source for most of the other platforms.
Despite Twitter's lower weights for alternative news URLs, it actually has a greater influence on alternative than mainstream news URLs, in terms of percentage of events caused, on all the other platforms/subreddits.
This is due to the fact that, even though it has lower weights, the largest proportion of alternative URL events are on Twitter.
After Twitter, The\_Donald and \dspol also have a strong influence on the alternative news URLs that get posted on other platforms.  The\_Donald has a stronger effect for alternative news URLs on all platforms except Twitter---although it still
has the largest alternative influence on Twitter, causing an estimated 2.72\% of alternative news URLs tweeted.
Interestingly, The\_Donald causes 8\% of \dspol's alternative news URLs, while \dspol's influence on The\_Donald is less, at 5.7\%.
For the mainstream news URLs the strength of influence is reversed.
Specifically, \dspol's influence on The\_Donald is 8.61\% whereas The\_Donald's influence on \dspol is 6.13\%.

In descending order, the influences on Twitter for mainstream news URLs are politics (4.29\%), \dspol (3.01\%), The\_Donald (2.97\%), worldnews (2.74\%), news (1.81\%), AskReddit (1.34\%), and conspiracy (1.04\%).
The strongest influences for alternative news URLs are, unsurprisingly, The\_Donald (2.72\%) and \dspol (1.96\%), followed by politics (1.10\%), worldnews (0.60\%), AskReddit (0.55\%), news (0.50\%), and conspiracy (0.46\%).  Twitter influences the alternative news URLs on other platforms to a large degree---but the largest alternative URL inputs to Twitter are The\_Donald and \dspol.
While we are only looking at a closed system of 8 different platforms and subreddits, we note that Twitter is undoubtedly effective at propagating information.
Thus the influence these two communities have on Twitter is likely to have a disproportional impact on the greater Web compared to their relatively minuscule userbase.

\subsection{Take-Aways}
In summary, the main take-aways from our influence analysis include: 
  1) Hawkes processes allow us to obtain a quantifiable influence between Web communities while taking into account influence from external sources of information; and 
  2) Twitter is influenced by smaller fringe Web communities, e.g., The\_Donald and \dspol  are responsible for around 6\% of mainstream news URLs and over 4.5\% of alternative news URLs posted to Twitter that appear on all three platforms.
  Considering Twitter's relative size, the impact of these fringe communities cannot be overstated.

\section{Related Work}
\label{sec:related_work}
We now review prior work on disinformation propagation dynamics in social networks
and on detecting false information sources.

\descr{Disinformation dynamics.} Kwon et al.~\cite{kwon2013prominent} study the characteristics of rumor propagation on Twitter.
They analyze a corpus of 1.7B tweets, covering three and a half years of Twitter data, and extract 104 viral events, which are
then annotated by human coders. They study debunked stories from false information busting sites such as Snopes and compare
temporal, linguistics, and structural characteristics to legitimate information.
Friggeri et al.~\cite{friggeri2014rumor} also use stories that Snopes determined as false to study the propagation and
the evolution of false information on Facebook, finding it to be quite bursty, and that posts containing a comment with a link
to Snopes are more likely to get deleted.
Kumar et al.~\cite{kumar2016disinformation} study the presence of hoaxes in Wikipedia articles. They report that while most
are detected quickly and have little impact, some end up cited widely on the Web.

Shao et al.~\cite{shao2016hoaxy} introduce Hoaxy, a platform providing information about the dynamics of false
information propagation on Twitter.
They also study a sample of 1.4M tweets, %
finding that the diffusion of fact-checking content lags that of false information by 10-20 hours and that
the top 1\% users with the most tweets share a much higher ratio of false information.
Finn et al.~\cite{finn2014investigating} present TwitterTrails, a website allowing users to study propagation
of false information on Twitter, i.e., to visualize indications of bursty activity, community
skepticism, temporal characteristics of propagation, as well as re-tweets networks. Also, Del Vicario et al.~\cite{del2016spreading} analyze how Facebook users perceive and react to conspiracy theories vs.~scientific stories, finding two polarized and homogeneous
communities that have similar content consumption patterns but exhibit different cascade dynamics.

Situngkir~\cite{situngkir2011spread} empirically studies an Indonesian hoax on Twitter,
finding that it spread broadly and  quickly (within two hours), and that it would have spread more
if a conventional media outlet did not publicly deny it.
He also argues that hoaxes can propagate easily if there is collaboration between the recipients of the hoax.
Arif et al.~\cite{arif2016information} present a case study based on a hostage crisis in Sydney, analyzing
5.4M tweets from three main perspectives:
1) volume (i.e., number of rumor-related messages per time interval), 2) exposure (i.e., number of individuals exposed to the rumor),
and 3) content production (i.e., whether the content is written by the user or is re-shared).
Andrews et al.~\cite{andrews2016keeping} study two crisis-related incidents on Twitter aiming to determine the effect of ``official''  accounts with respect to the containment of rumors. Authors show that official account can significantly contribute to stopping the propagation of
the rumor by actively engaging in conversations related to the incidents.
Finally, Mendoza et al.~\cite{mendoza2010twitter} study the  dissemination of false rumors vs.~confirmed news on Twitter the days
following the 2010 earthquake in Chile, concluding that an aggregate analysis on the flow of tweets can distinguish
the former from the latter.

\descr{Detecting false information sources.}
Shah et al.~\cite{shah2011rumors} formulate the problem of finding the source of false information as a maximum likelihood estimation
problem, using a metric called rumor centrality.
They evaluate it for all nodes in the network using a simple linear time message-passing algorithm, and the node with the highest rumor
centrality is deemed to be the most likely source.
Authors show, experimentally, that the model can distinguish the source of false information with a maximum error of 7 hops for general
networks and 4  for tree networks.
Wang et al.~\cite{wang2014rumor} study the problem from a statistical point of view, proposing a source detection
framework, also based on rumor centrality, which supports multiple snapshots of the network during the
false information spread. They show that using two network snapshots instead of one can significantly improve detection.

Budak et al.~\cite{budak2011limiting}  study the notion of competing campaigns in a social network and address the problem of
influence limitation to counteract the effect of misinformation.
Nguyen et al.~\cite{nguyen2012sources} look for the $k$ users that are most suspected to have originated false information, using
a reverse diffusion process along with a ranking process.
Seo et al.~\cite{seo2012identifying} aim to identify the source of rumors in online social networks by injecting monitoring nodes
across the social graph. They propose an algorithm that observes the information received by the monitoring nodes in order to identify the source.
They indicate that with sufficient number of monitoring sources they can recognize the source with high accuracy.

Finally, Starbird~\cite{starbird2017examining} performs a qualitative analysis on tweets pertaining to shooting events and
conspiracy theories, using graph analysis on the domains linked from the tweets, and provides insight on how various websites work
to promote conspiracy theories and push political agendas.

\descr{Remarks.} In contrast to prior work, this paper provides insights on disinformation dynamics on social networks from a
comprehensive (i.e., multi-service) point of view. In other words, we study major online social networks and various news sites that actively
contribute to information diffusion across the Web, specifically, analyzing the dynamics and information flow of
Reddit, 4chan, Twitter, and several news sites.
Furthermore, we measure the influence that each platform has to each other using a rigorous statistical model---namely, Hawkes processes.

\section{Discussion \& Conclusion}
\label{sec:conclusion}

This paper explored how mainstream and fringe Web communities share mainstream and alternative news sources with a particular focus on how communities influence each other.
We collected millions of posts from Twitter, Reddit, and 4chan, and analyzed the occurrence and temporal dynamics of news shared from 45 mainstream and 54 alternative news sites.
We found that users on the different platforms prefer different news sources, especially when it comes to alternative ones.
We also explored complex temporal dynamics and we discovered, for example, that Twitter and Reddit users tend to post the same stories within a relatively short period of time, with 4chan posts lagging behind both of them.
However, when a story becomes popular after a day or two, it is usually the case it was posted on 4chan first, lending some credence to 4chan's supposed influence on the Web.

Using Hawkes processes, we also modeled the influence the individual platforms have on each other, while also taking into account influence that comes from external sources of information.
We found that the interplay between platforms manifests in subtle, yet meaningful ways.
For example, of all the platforms and subreddits, Twitter by far has the most influence in terms of the number of URLs it causes to be posted to other platforms, and contributes to the share of alternative news URLs on the other platforms to a much greater degree than to the share of mainstream news URLs.
After Twitter, The\_Donald subreddit and \dspol are the next most influential when it comes to alternative news URLs.
For such URLs, The\_Donald is less influenced by the other platforms than \dspol, and has a higher background rate, i.e., more of the URLs posted there come from other sources.

To the best of our knowledge, our analysis constitutes the first attempt to characterize the dissemination of mainstream and alternative news across multiple social media platforms, and to estimate a quantifiable influence between them.
Overall, our findings shed light on how Web communities influence each other and can be extremely useful to better understand and detect false information as well as informing the design of systems that aim to trace the origins of fake stories and mitigate their dissemination.

\descr{Limitations.}
This work presented an initial attempt to measure the influence among popular Web communities.
To this end, we considered news propagation on Web communities as dictated by the posting of URLs from mainstream and alternative news sources.
Naturally, this approach is not without limitations, as we did not account for the content of the information, which can exist and propagate in various forms, such as textual claims, images, and videos.
For example, there may be a lot of direct information transferred from 4chan occurring via screenshots, due to its ephemeral nature and orientation towards images.
In fact, this highlights the complexity of the problem, as, e.g., distinguishing if an image delivers the same information as a textual claim in an article, is far from a straightforward task.
Moreover, we did not examine the content of the news stories shared.

\descr{Future Work.} As part of future work, we plan to explore advanced image recognition techniques to look for screenshots shared among the different platforms, as well as Natural Language Processing methods to determine whether stories become a part of the platform's narrative of events -- i.e., whether users continue to talk about stories without actually posting a relevant URL itself.
We believe efforts into understanding how the growing phenomenon of alternative and fake information sources affects multiple platforms can help inform detection and mitigation techniques against misinformation and disinformation campaigns, and our work constitutes a first step in that direction.

\descr{Acknowledgments.} This project has received funding from the European Union's Horizon 2020 Research and Innovation program under the Marie Sk\l{}odowska-Curie ENCASE project (Grant Agreement No. 691025).
This work reflects only the authors' views and the Agency and the Commission are not responsible for any use that may be made of the information it contains.

\bibliographystyle{abbrv}

\end{document}